\documentclass[aps,pra,reprint,showpacs,amsmath,amssymb]{revtex4-1}
\pdfoutput=1
\usepackage{bm}
\usepackage{hyperref}
\usepackage{color}
\usepackage{overpic}
\usepackage{mathtools}
\usepackage{bbold}
\usepackage{graphicx}
\graphicspath{{./figures/}}
\newcommand{\dd}{\text{d}}
\newcommand{\ee}{\text{e}}
\newcommand{\ii}{\text{i}}

\newcommand{\ket}[1]{\left| #1 \right\rangle}
\newcommand{\braket}[2]{\left\langle #1 \middle| #2 \right\rangle}
\renewcommand{\vec}[1]{\bm{#1}}

\begin{document}
\title{Quantum Rotor Theory of Systems of Spin-2 Bosons}
\author{Matja\v{z} Payrits}
\author{Ryan Barnett}
\affiliation{Department of Mathematics, Imperial College London,
London SW7 2AZ, United Kingdom}
\begin{abstract}
We consider quantum phases of tightly-confined spin-2 bosons in an external
field under the presence of rotationally-invariant interactions.  Generalizing
previous treatments,  we show how this system can be mapped onto a quantum
rotor model.  Within the rotor framework, low-energy excitations about
fragmented states, which cannot be accessed within standard Bogoliubov theory,
can be obtained.  In the spatially extended system in the thermodynamic limit
there exists a mean-field ground state degeneracy between a family of nematic
states for appropriate interaction parameters. It has been established that
quantum fluctuations lift this degeneracy through the mechanism of
order-by-disorder and select either a uniaxial or square-biaxial ground state.
On the other hand, in the full quantum treatment of the analogous
single-spatial mode problem with finite particle number it is known that, due
to symmetry restoring fluctuations, there is a unique ground state across the
entire nematic region of the phase diagram.  Within the established rotor
framework we investigate the possible quantum phases under the presence of a
quadratic Zeeman field, a problem which has previously received little
attention. By investigating wave function overlaps we do not find any
signatures of the order-by-disorder phenomenon which is present in the
continuum case. Motivated by this we consider an alternative external potential
which breaks less symmetry than the quadratic Zeeman field. For this case we do
find the phenomenon of order-by-disorder in the fully quantum system. This is
established within the rotor framework and with exact diagonalization.
\end{abstract}
\maketitle

\section{Introduction} Ultracold spinor atoms provide  simple and
experimentally well-controlled many-body systems with internal degrees of
freedom. The simplest case is that of a spin-1 gas.   The atomic spin-1 species
${}^{23}\text{Na}$ and ${}^{87}\text{Rb}$ have been the subject of numerous
experiments (see for example the reviews in
\cite{Stamper-Kurn2013,Kawaguchi2012} and references therein), prompting
detailed theoretical investigation of the exact spectra for tightly confined
spin-1 atoms
\cite{Law1998,Ho2000,Koashi2000,Mueller2006,Barnett2010,Barnett2011,
Lamacraft2010}  as well as their mean-field properties in the thermodynamic
limit \cite{Ho1998,Ohmi1998}.  The higher spin-2 hyperfine multiplet of
${}^{87}\text{Rb}$ was found to be stable and amenable to experimental
manipulation \cite{Chang2004,Kuwamoto2004,Schmaljohann2004,Burke1997},
prompting the development of a number of spin-2 exact and mean-field
theoretical results \cite{Ciobanu2000,Koashi2000,Ueda2002,Song2007,Turner2007}.
In particular, Gross-Pitaevskii mean field theory has been used to describe
spin-2 condensates under linear and quadratic Zeeman fields
\cite{Saito2005,Uchino2010}.  Exact quantum results have also been found for
spin-2 systems under tight spatial confinement under no external fields, which
can be simply extended to results with a linear Zeeman field by a gauge
transformation \cite{Ueda2002,Koashi2000}.  The case of the quantum states of
tightly confined bosons under the presence of a quadratic Zeeman field,
however, has remained less well understood.

Spinor condensates also provide a convenient platform with which to study the
phenomenon of order-by-disorder \cite{Villain1980,Henley1987,Henley1989}, or
the selection of a particular mean-field ground state from a set of
accidentally degenerate ones due to fluctuations. This subject has
traditionally been of importance in elucidating the ground-state structure of
frustrated magnetic systems but has also been investigated as a relevant mechanism
in several cold-atom systems
\cite{Wessel2005,Song2007,Turner2007,Wu2008,Zhao2008,Barnett2012,You2012,Zheng2013,Sun2014,Payrits2014}.
These offer the possibility of experimentally realizing a phenomenon whose
observation is still contentious in magnetic systems \cite{Ross2014,Petit2014}.
In spin-2 systems, order by disorder has been predicted to determine the ground
state in the absence of a quadratic Zeeman field for species with scattering
lengths within a certain range, termed the nematic region. Furthermore, the
mechanism has been predicted to introduce a first-order phase transition
between two parts of the nematic region in which fluctuations select different
members of the accidentally degenerate family \cite{Song2007,Turner2007}.

One of the goals of the present article is to present analytical results
for tightly confined spin-2 atoms in the presence of a quadratic Zeeman field.
The results are obtained by utilizing an exact mapping of the interacting
many-body system Hamiltonian to a 5-dimensional quantum rotor Hamiltonian,
i.e., that of a single particle moving on the 4-sphere. Similar mappings have
previously been employed to study the double-well problem \cite{Anglin2001},
dipolar condensates \cite{Armaitis2013} and particularly the analogous tightly
bound spin-1 problem \cite{Barnett2010,Barnett2011,Jing2011,Buchmann2013}.

One of the main perceived advantages of the rotor methodology is that it allows
one to treat excitations about fragmented states
\cite{Penrose1956,Nozieres1982}.  Applying the Penrose-Onsager criterion for
Bose-Einstein condensation, fragmented condensates are defined as those whose
reduced single-particle density matrix has more than one extensive eigenvalue.
When there are exactly two such eigenvalues, one may envision the state as a
condensate of particle pairs. One encounters such a case in the spin-2 problem
in the presence of a large negative quadratic Zeeman field. Such a state cannot
be approximated by a coherent state, invalidating the use of Bogoliubov theory,
typically the first line of attack in calculating excitation spectra about
non-fragmented condensates. The rotor mapping, on the other hand, suffers no
pathologies in the fragmented case and provides simple analytical expressions
for the excitation spectra.

Previous results by Koashi and Ueda \cite{Koashi2000,Ueda2002} indicate that
the exact quantum results in the absence of a quadratic Zeeman field do not
mirror the continuous accidental degeneracy of the mean-field analysis. Rather,
the ground state is non-degenerate and the same across the entire nematic
region and no traces of the order-by-disorder induced phase transition are
manifest.  One may hope that the signature of the transition could nevertheless
be observed in the magnetic response to the quadratic Zeeman field, which
serves to classically  orient the nematic order parameter. In the present
article we find through an evaluation of the wave function overlaps obtained
through the rotor mapping, that such a signature is nevertheless not present.  

Motivated by this, we have applied the rotor mapping to the analysis of an
alternative potential that  \emph{does not}  fully break the mean-field
degeneracy, noting that a large class of quadratic potentials can be
experimentally obtained with external microwave fields.  Using the rotor
framework we show that including quantum corrections select a unique ground
state.   Furthermore, this selection is explicitly demonstrated through
an exact diagonalization numerical approach involving a modest number of atoms.
Importantly, we find that the overlap of the obtained ground states
with any mean-field state tends to zero with increasing particle number, a
stark departure from the standard mean-field states obtained in the
continuum. On the other hand, the ordering is apparent in the spin-component
occupation numbers, which can be readily experimentally probed.

This article is organized as follows. In Sec.~\ref{sec:background} we first
describe the continuum Hamiltonian for a spin-2 cold Bose gas. Mean field
results and a phase diagram dependening on the Zeeman field and the differences
in distinct total-spin scattering lengths, considered tunable, are presented.
From thereon we focus on the \emph{nematic} region of the phase diagram. In
Sec.~\ref{sec:sma-and-exact-diag} we next consider the single mode
approximation, relevant to a tight trap in which spatial degrees of freedom are
taken to be completely frozen out, yielding an effectively zero-dimensional
Hamiltonian.  Exact results on this Hamiltonian in the absence of a quadratic
Zeeman field are summarized in Subsec.~\ref{sec:known-quantum-pd-aspects}. In
the following subsection we briefly outline the exact-diagonalization method
used to obtain numerical results in the present work. In
Sec.~\ref{sec:rotor-mapping} the spin-2 rotor mapping is introduced. Though
having a real spectrum, the Hamiltonian thus obtained is in general
non-Hermitian. The Hermitianizing transform is generally difficult to find, but
feasible in special cases.  Subsec.~\ref{sec:special-case-hermitianizing}
considers such a case when one of the Hamiltonian parameters is zero.
Section~\ref{sec:large-q-limits} considers general parameter configurations in
the presence of a large quadratic Zeeman field. Large positive and negative
values are considered in separate subsections. In both cases a simple
approximate Hermitianizing transform may be found, leading to effective
harmonic oscillator Hamiltonians. In subsection~\ref{sec:q-overlaps} we further
present analytical expressions for overlaps with the relevant mean-field states
for both cases and compare them to numerics.

Finally, in Sec.~\ref{sec:obd} we discuss the disparity in qualitative
nematic-region behaviour arising from the mean-field and full quantum
treatments in the presence of a quadratic Zeeman term. The latter contains no
visible traces of order-by-disorder that is manifest in the former.  Motivated
by this, we introduce an alternative external potential which does not break
the mean-field degeneracy.  We show that beyond-mean-field corrections in this
system select unique ground states, and therefore interpret the phenomenon as
order by disorder.  We derive analytical expressions for certain aspects of
this state and numerically assess them.

\section{Background}
\label{sec:background}
\subsection{Spinor Hamiltonian}

We begin by describing the Hamiltonian governing the underlying physical
system, a collection of cold interacting spin-2 bosons in a scalar trapping
potential \footnote{By a scalar potential we mean one that couples to all
magnetic sublevels approximately equally, such as the potential of an optical
trap and unlike that of a magnetic trap.} and a magnetic field, manifesting
itself through a linear and a quadratic Zeeman term. The full first-quantized
Hamiltonian is
\begin{eqnarray}
	\hat{H}_\text{1st} &=& \sum_{i}^N \hat{H}_i^{(1)} + \sum_{i < j}
	\hat{V}^{(2)}_{i,j} \qquad \text{with}\nonumber\\
	\hat{H}_i^{(1)} &=& \frac{1}{2m} \hat{p}_i^2 + V(\hat{\vec{r}}_i)
	+ p \hat{F}^z_i + q (\hat{F}^z_i)^2.
	\label{eq:first-quant-hamiltonian}
\end{eqnarray}
Here $N$ is the total particle number, $m$ the atomic mass, $V$ the external
potential and $\hat{p}_i$, $\hat{\vec{r}}_i$ and $\hat{F}_i^z$ the $i$-th
particle's momentum, position and $z$-component of spin operators,
respectively. $p$ and $q$ are the linear and quadratic Zeeman coefficients,
respectively.

As detailed in many standard resources, such as \cite{Pethick2008}, the
interparticle potential $\hat{V}^{(2)}_{i,j}$ between the $i$-th and $j$-th
particles is short-range and dominated by the s-wave component, i.e., it
depends predominantly on the distance between the atoms. It is  well
approximated by a delta function with a prefactor proportional to the
scattering length.  There are three different scattering lengths $a_S$ for the
three distinct values of the total spin of a pair of particles $S$ allowed by
interchange symmetry, that is, 0, 2 and 4.  The potential of a pair of
particles can thus be written as \cite{Stamper-Kurn2013,Kawaguchi2012}
\begin{equation}
	\hat{V}^{(2)}_{i,j} = \frac{1}{2} \delta\left( \hat{\vec{r}}_i -
	\hat{\vec{r}}_j \right) \left( c_0 \hat{\mathbb{1}}_i \otimes
	\hat{\mathbb{1}}_j + c_1 \hat{\vec{F}}_i \cdot \hat{\vec{F}}_j + 5 c_2
	\hat{P}^{(0)}_{i,j} \right)
	\label{eq:pair-interaction}
\end{equation}
where $\hat{P}^{(0)}_{i,j}$ is the projection operator onto the spin singlet
state of the pair, $\hat{\vec{r}}_i$ and $\hat{\vec{F}}_i$ are the $i$-th
particle's position and spin operator, respectively. Employing units with
$\hbar = 1$, used hereafter, the $c$ constants may be expressed in terms of
scattering lengths as
\begin{align}
c_0 &= \frac{4 \pi}{7m}   \left( 4 a_2
	+ 3 a_4 \right)  \notag  \\ 
	c_1 &= \frac{4 \pi}{7m} \left( a_4 - a_2 \right)  \label{eq:c-constants} \\
	c_2 &= \frac{4 \pi}{5m} \left(
		a_0 - a_4 \right) + \frac{8 \pi}{7m}\left( a_4 - a_2 \right) \notag			
\end{align}

Second-quantizing Hamiltonian~(\ref{eq:first-quant-hamiltonian}) above yields:
\begin{subequations}
\label{eq:full-second-quantized-hamilt}
\begin{eqnarray}
	\hat{H}_\text{2nd} &=& \int \dd^3 \vec{r}\, \left( \hat{\mathcal{H}}_0(\vec{r})
	+ \hat{\mathcal{H}}_q(\vec{r}) +
	\hat{\mathcal{H}}_I(\vec{r})\right)\label{eq:3d-hamiltonian}\\
	\hat{\mathcal{H}}_0 &=& \hat{\psi}^\dagger_\alpha \left( -\frac{1}{2m}
	\nabla^2 + V(\vec{r}) \right) \hat{\psi}_\alpha\label{eq:second-quant-hamilt-zero}\\
	\hat{\mathcal{H}}_q &=& p \hat{\mathcal{F}}^z + q \hat{\mathcal{Z}} \quad
	\text{with}\nonumber\\
	\hat{\mathcal{F}}^i &=& \hat{\psi}^\dagger_\alpha F^i_{\alpha \beta} \hat{\psi}_\beta
	\quad \text{and} \quad \hat{\mathcal{Z}} = \hat{\psi}^\dagger_\alpha
	(F^z)^2_{\alpha \beta} \hat{\psi}_\beta\label{eq:second-quant-hamilt-q}\\
	\hat{\mathcal{H}}_I &=& : \frac{c_0}{2} \hat{n}^2 + \frac{c_1}{2}  \hat{\vec{\mathcal{F}}}^2 :
	+ \,\frac{c_2}{2}  \hat{\mathcal{A}}^\dagger \hat{\mathcal{A}} \quad
	\text{with}\nonumber\\
	\hat{n} &=& \hat{\psi}^\dagger_\alpha \hat{\psi}_\alpha \quad
	\text{and} \quad \hat{\mathcal{A}} = \sum_{\alpha = -2}^2 (-1)^\alpha
	\hat{\psi}_\alpha \hat{\psi}_{-\alpha}\label{eq:second-quant-hamilt-interaction}
\end{eqnarray}
\end{subequations}
where $\hat{\psi}_\alpha (\vec{r})$ are the annihilation operators for bosons
in the $m = \alpha$ magnetic sublevel at $\vec{r}$. $\hat{\mathcal{F}}^i
(\vec{r})$ stands for the $i$-th component of the total spin density operator
whereas  $F^i$ is the $i$-th spin-2 matrix. The positional dependence of
creation/annihilation operators and densities has been suppressed above for
brevity. Except for the definition of $\hat{\mathcal{A}}$, repeating indices
imply the Einstein summation convention.  The colon delimiters represent normal
ordering. Note that $\hat{\mathcal{H}}_0$ and $\hat{\mathcal{H}}_q$ are the
second-quantized forms of the kinetic/potential and linear/quadratic Zeeman
terms of the single particle part in
Hamiltonian~(\ref{eq:first-quant-hamiltonian}) whereas $\hat{\mathcal{H}}_I$ is
the second-quantized form of the two-particle interaction in
Eq.~(\ref{eq:pair-interaction}).  It is also worth noting that the operator
$\hat{\mathcal{A}}$ may be loosely interpreted as an annihilation operator for
a spin-singlet pair of bosons \cite{Koashi2000,Ueda2002}.

\subsection{Mean-field phase diagram at $q = 0$}

\begin{figure}
	\includegraphics[width=\columnwidth]{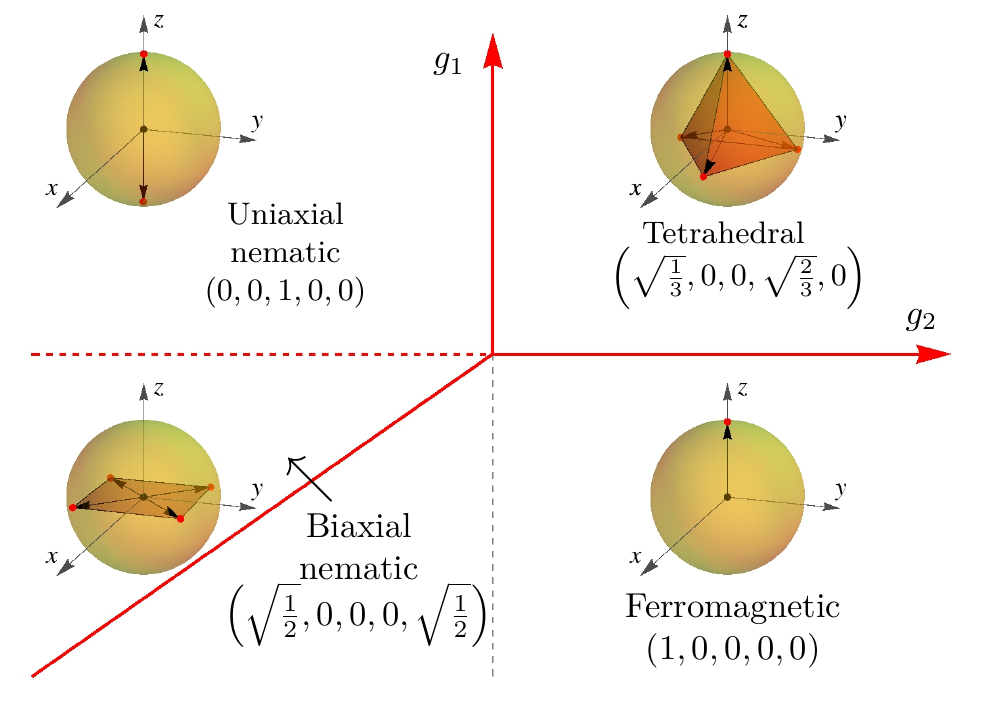}
	\caption{Mean field phase diagram at $q=0$.}
	\label{fig:mf-phase-diagram}
\end{figure}

In determining ground states in the  thermodynamic limit we may invoke
Gross-Pitaevskii mean field theory which consists of replacing field operators
with their expectation values, $\hat{\psi}_\alpha(\vec{r}) \rightarrow
\psi_\alpha (\vec{r}) \equiv \langle \hat{\psi}_\alpha (\vec{r}) \rangle$. In
the continuum, i.e., zero external potential, we may Fourier transform the
operators to $\hat{\psi}_{\vec{p}, \alpha}$ and for the ground states further
consider only the zero momentum, $\vec{p} = \vec{0}$ components.  This reduces
classifying the various phases to describing the five-component order parameter
$\chi_\alpha \equiv \langle \hat{\psi}_{\vec{0},\alpha} \rangle/\sqrt{N}$ where
$N$ is the total particle number, up to rotational and $\text{U}(1)$ phase
symmetries. These order parameters are shown in Fig.~\ref{fig:mf-phase-diagram} with
respect to $c_{1,2}$.  Also shown are the Majorana representations of the order
parameters, which show the rotational symmetries of the states
\cite{Barnett2006a,Stamper-Kurn2013}.

In the nematic region, where $g_2 < \min(0, 4 g_1)$, there is an additional
continuous accidental degeneracy of states with rectangular Majorana
representations. A general nematic state's order parameter can be written,
up to the aforementioned symmetries, as
\begin{equation}
	\vec{\chi}_\text{n}(\eta) = \left( \frac{\text{sin}\,\eta}{\sqrt{2}},
	0, \text{cos}\, \eta, 0, \frac{\text{sin}\,\eta}{\sqrt{2}} \right)
	\label{eq:degenerate-nematic-states}
\end{equation}
where $\eta$ parametrizes the degeneracy. The two states shown in
Fig.~\ref{fig:mf-phase-diagram} are representatives of higher symmetry,
obtained by setting $\eta = 0\: (\eta = \pi/2)$, which are referred to as
uni(bi)-axial states. While all $\eta$ states are degenerate at the mean-field
level, fluctuations  lift the degeneracy through the phenomenon of
order-by-disorder, selecting the uni(bi)-axial state for $c_1 > 0\; (c_1 < 0)$
\cite{Song2007,Turner2007}.  This leads to the dotted phase boundary in
Fig.~\ref{fig:mf-phase-diagram}.

Mean-field behavior in the thermodynamic limit has also been extensively
investigated at non-zero $q$ in \cite{Uchino2010} leading to a number of new
phases, where some of the phase boundaries with respect to $c_{1,2}$ and $q$
had to be numerically determined. While the $q=0$ ground states were invariant
to changes in $c_{1,2}$ that remained within the same phase \footnote{The
term \emph{inert} is sometimes used to describe such states
\cite{Stamper-Kurn2013}.},  ground states in some of the new $q \neq 0$ phases
vary continuously with the parameters. Lastly and perhaps most importantly, the
nematic accidental degeneracy in $\eta$ is lifted for any $q \neq 0$.

\section{Single mode approximation and exact diagonalization}
\label{sec:sma-and-exact-diag}
\subsection{Single mode Hamiltonian}
\label{sec:sma-hamiltonian}
In the remainder of the text we consider a tightly bound many-body system with
a fixed number of particles $N$. This implies we can write
$\hat{\psi}_\alpha(\vec{r})$ from Eq.~(\ref{eq:full-second-quantized-hamilt})
as $\hat{\psi}_\alpha(\vec{r}) = \phi_0(\vec{r}) \hat{a}_\alpha$ where
$\phi_0(\vec{r})$ is the unit-normalized lowest spatial mode of the system and
$\hat{a}_\alpha$ is the annihilation operator for a boson in this lowest
spatial mode with magnetic number $m = \alpha$, since the tightness of the
confining potential energetically prohibits spatially excited states. This is
traditionally called the single mode approximation or SMA \cite{Law1998}. For
convenience we also define the vector of operators $\hat{\vec{a}} = \left(
\hat{a}_2, \hat{a}_1, \hat{a}_0, \hat{a}_{-1}, \hat{a}_{-2} \right)^T$.
Substituting the $\hat{\psi}_\alpha$ as above and integrating out the spatial
components of the Hamiltonian  Eq.~(\ref{eq:full-second-quantized-hamilt})
yields
\begin{equation}
	\label{eq:many-body-hamiltonian}
	\hat{H}_\text{SMA} = \frac{g_1}{2 N}\hat{\vec{F}}^2 + \frac{g_2}{2
	N}\hat{A}^\dagger\hat{A} + q \hat{Z}.
\end{equation}
plus constants. Here $g_i = n_0 c_i$ where $n_0 = N \int \dd\vec{r}\,
\left|\phi(\vec{r})\right|^4$ and $c_i$ are
defined in Eq.~(\ref{eq:c-constants}). The uppercase operators are obtained
from their calligraphic density counterparts in
Eq.~(\ref{eq:full-second-quantized-hamilt}) by letting $\hat{\psi}_\alpha
\rightarrow \hat{a}_\alpha$, e.g., $\hat{F}^i = \hat{a}^\dagger_\alpha
F^i_{\alpha \beta} \hat{a}_\beta = \hat{\vec{a}}^\dagger F^i \hat{\vec{a}}$
where $F^i$ still represents the $i$-th spin matrix.

The Hamiltonians~(\ref{eq:full-second-quantized-hamilt})
and~(\ref{eq:many-body-hamiltonian}) evidently conserve total particle number
$\hat{N}$ and, as noted above, we consider it fixed at $N$. This allows us to
drop terms arising from the spatial integrals of $\hat{\mathcal{H}}_0$ and
$:\hat{n}^2:$ of Eq.~(\ref{eq:full-second-quantized-hamilt}) and to simplify
the contribution of $:\mathrel{\hat{\vec{\mathcal{F}}}^2}:\: = \hat{\vec{\mathcal{F}}}^2 -
C \hat{n}$ to $\hat{\vec{F}}^2$. Hamiltonian~(\ref{eq:many-body-hamiltonian})
further commutes with $\hat{F}^z$ and can thus be simultaneously diagonalized.
In the remainder of this text we consider fixed $\hat{F}^z$ eigenspaces, most
often the nullspace, allowing us to drop the linear Zeeman term as in
Eq.~(\ref{eq:many-body-hamiltonian}).

\subsection{Known aspects of the quantum phase diagram}
\label{sec:known-quantum-pd-aspects}
At zero quadratic Zeeman field $q$, the exact spectrum of the tightly confined
system is known \cite{Ueda2002}. Potentially degenerate eigenlevels can be
labelled by the set $\left\{ N_0, N_S, F, F_z \right\}$ where $F_z$ is the
eigenvalue of $\hat{F}_z$ and $F$ is such that the eigenvalue of
$\hat{\vec{F}}^2$ equals $F(F+1)$. $N_S$ can be interpreted as the number of
spin-singlet pairs and $N_0 \equiv N - 2 N_S$ as the number of bosons not in
the singlet state. As mentioned above, this analogy is only a loose one, as
$\hat{A}$ and $\hat{A}^\dagger$ do not obey Bosonic commutation relations.
However, the commutation relations of these and a third operator, which the
authors of \cite{Ueda2002} denote by $\hat{\mathcal{S}}_z \equiv
\frac{1}{4}(2 \hat{N} + 5)$, can be seen to be those of the Lie algebra
$\mathfrak{su}(1,1)$, closely related to that of $\mathfrak{su}(2)$, the spin
algebra.  This allows for an elegant derivation of the joint $\hat{A}^\dagger
\hat{A}$ and $\hat{\mathcal{S}}_z$ eigenstates in analogy with the raising and
lowering operator approach to the spin algebra. Technically, $N_S$ and $N_0$
are defined such that the eigenvalue of $\hat{A}^\dagger \hat{A}$ equals
$\left( N_0 + \frac{1}{2} \right) \left( N_0 + \frac{5}{2} \right)$ and $N_0 +
2 N_S = N$. In terms of the above quantum numbers, the energies are given by
\begin{equation}
	E = \frac{g_1}{2} \left[ \frac{F}{N}\left( F + 1 \right) - 6 \right] +
	g_2 \frac{N_S}{N} \left( N + N_0 + 3 \right)
	\label{eq:zero-q-exact-energies}
\end{equation}
The easily obtained ground states show interesting parallels with the
mean-field phase diagram. In the ferromagnetic region, the ground state $N_S$
is zero and $F = 2 N$ is maximized, while in the tetrahedral
region the ground state $N_S$ and $F$ are both zero.

The nematic-region ground state is, however, less easily reconciled with
its mean-field counterparts, as the ground state is non-degenerate and unique
across the entire nematic region. It consists only of singlet pairs and
potentially a singlet trio, maximizing $N_S$ and minimizing $F$.

On the other hand, the case where $q \neq 0$ is much less well-understood
analytically as $N_S$ or $N_0$ are no longer good quantum numbers.  We shall
focus on this regime in the following.  Analytical results are obtained via the
rotor mapping and contrasted with the numerical results obtained through exact
diagonalization, which we briefly describe next.

\subsection{Exact diagonalization}

Due to the effective spatial 0-dimensionality of our tightly bound system our
problem is that of diagonalizing a five-mode many-body Hamiltonian.  Further
fixing $N$ and $F_z$, the relevant Fock bases may be enumerated by three
independent occupation numbers. The sizes of the bases hence scale as $N^3$
with particle number $N$ making it quite feasible to diagonalize
Hamiltonian~(\ref{eq:many-body-hamiltonian}), or at least find the ground state
and its energy, at fixed values of $g_{1,2}$, $q$, $F_z$ and $N$ with regular
desktop hardware in timescales on the order of hours for up to about 300
particles.

Denoting Fock states by
\begin{equation}
	\ket{n_2, n_1, n_0, n_{-1}, n_{-2}} \equiv \prod_{m = -2}^2
	\frac{\hat{a}_m^{\dagger\,n_m}}{\sqrt{n_m !}} \ket{0},
	\label{eq:fock-state}
\end{equation}
one way of enumerating the entire Fock basis for fixed $N$ and $F_z$ is by
considering $n_2, n_1$ and $n_{-2}$ as independent variables and letting $n_0 =
N + F_z - 3 n_2 - 2 n_1 + n_{-2}$ and $n_{-1} = 2 n_2 + n_1 - 2 n_{-2} - F_z$.
The ranges of the independent $n$ variables are cumbersome to state but can
easily be found programmatically. What remains is expressing the terms of
Hamiltonian~(\ref{eq:many-body-hamiltonian}) with respect to this basis and
diagonalizing the resulting sparse matrices, which can be accomplished with
standard numerical packages.

\section{The rotor mapping}
\label{sec:rotor-mapping}

Here we present the primary calculational tool allowing the derivation of
analytical results of the present text, the rotor mapping. This has been
introduced for spin-1 in \cite{Barnett2010} and expanded upon in
\cite{Barnett2011}. The latter reference also includes a brief discussion of
the mapping for spin-2 systems in the absence of external fields.  In this
section we review the key points of the mapping in general and extend it to
include an external potential for the spin-2 case.  In
subsection~\ref{sec:cartesian-basis} we first comment on the natural basis of
the spin-2 representation for use with the mapping and state the form of the
spin matrices in it. In subsection~\ref{sec:rotor-hamiltonian} we briefly
review the main steps of the mapping and derive the rotor Hamiltonian for our
system. We comment on its properties in different regions of the parameter
space, particularly Hermiticity.
Subsection~\ref{sec:special-case-hermitianizing} outlines the process of
exactly Hermitianizing the Hamiltonian in the special case when $g_1 = 0$,
demonstrating the equivalence of two Hermitian Hamiltonians, the many-body
Hamiltonian~(\ref{eq:many-body-hamiltonian}) and that of a single particle on
the 4-sphere in a specific potential. This subsection also introduces some of
the methods employed in later sections to make approximate low-energy
Hamiltonians Hermitian.

\subsection{Cartesian basis}
\label{sec:cartesian-basis}

While the canonical spin-2 matrices are complex and Hermitian, there exists a
basis in which they are completely imaginary and antisymmetric. This allows one
to map the original spin operators onto linear combinations of generalized
angular momentum operators, significantly simplifying the analysis. The
underlying reason for this is that bosonic representations of the spin group
$\text{SU}(2)$ can also be thought of as representations of the real group of
rotations $\text{SO}(3)$.

There is a simple heuristic method of finding such a basis, based on
transforming from standard complex spherical harmonics to real ones. We state
results in terms of annihilation operators rather than the underlying
single-particle basis. Using the phase convention $Y_l^{-m}(\theta,\varphi) =
(-1)^m Y_l^m (\theta, \varphi)^* \propto \ee^{- \ii m \varphi}$, one arrives at
the  Cartesian annihilation operators:
\begin{eqnarray}
	\hat{b}_1 &=& \hat{a}_0\nonumber\\
	\begin{split}
		\hat{b}_2 &= \frac{\ii}{\sqrt{2}} \left( \hat{a}_1 + \hat{a}_{-1}
		\right)\\
		\hat{b}_3 &= \frac{1}{\sqrt{2}}\left( \hat{a}_{-1} - \hat{a}_1
		\right)
	\end{split} \;&&\;
	\begin{split}
		\hat{b}_4 &= \frac{\ii}{\sqrt{2}} \left( \hat{a}_{-2} -
		\hat{a}_2 \right)\\
		\hat{b}_5 &= \frac{1}{\sqrt{2}}\left( \hat{a}_2 + \hat{a}_{-2}
		\right)
	\end{split}
	\label{eq:b-operators}
\end{eqnarray}
Gathering these into $\hat{\vec{b}} = \left( \hat{b}_1, ..., \hat{b}_5
\right)^T$ we may neatly express parts of
Hamiltonian~(\ref{eq:many-body-hamiltonian}) in terms of the
$\hat{b}$-operators. The singlet operator becomes simply $\hat{\mathcal{A}} =
\hat{\vec{b}}\cdot\hat{\vec{b}}$. The spin and quadratic Zeeman operators
become $\hat{F}^i = \hat{\vec{b}}^\dagger M^i \hat{\vec{b}}$ and $\hat{Z} =
\hat{\vec{b}}^\dagger (M^z)^2 \hat{\vec{b}}$, where the $M^i$ are the pure
imaginary antisymmetric spin matrices expressed in the new single-particle
basis.

To state their forms concisely, let us introduce a family of simple
antisymmetric matrices. Let $S^{(-)}_{i j}$ be the 5-by-5 matrix with $1$ in the
$i$-th row and $j$-th column, $-1$ in the $j$-th row and $i$-th column, and $0$
elsewhere. That is, $\left( S^{(-)}_{i j} \right)_{\alpha \beta} = \delta_{i
\alpha} \delta_{j \beta} - \delta_{i \beta} \delta_{j \alpha}$. The $M^i$ are
then:
\begin{eqnarray}
	M^x &=& -\ii \left( \sqrt{3} \, S^{(-)}_{1 2} - S^{(-)}_{2 5} + S^{(-)}_{3 4} \right)\nonumber\\
	M^y &=& -\ii \left( \sqrt{3} \, S^{(-)}_{1 3} + S^{(-)}_{2 4} + S^{(-)}_{3 5} \right)\nonumber\\
	M^z &=& -\ii \left( S^{(-)}_{2 3} + 2 \, S^{(-)}_{4 5} \right)
	\label{eq:spin-matrices-transf}
\end{eqnarray}
For convenience also denote $Q \equiv (M^z)^2 = \text{diag}\,(0, 1, 1, 4, 4)$.

\subsection{The rotor Hamiltonian}
\label{sec:rotor-hamiltonian}

Next we construct the overcomplete basis \footnote{It is evident that the basis
is not orthogonal.  Showing completeness is relatively straightforward but
tedious.}
\begin{equation}
	\ket{\vec{\Omega}}=\frac{1}{\sqrt{N !}} \left( \vec{\Omega}\cdot
	\vec{\hat{b}}^\dagger \right)^N \ket{0}
	\label{eq:rotor-basis-definition}
\end{equation}
where $\vec{\Omega}$ is a norm-1 5-component \emph{real} vector, i.e.,
belonging to the 4-sphere $\mathcal{S}^4$.

It turns out that the elements of basis~(\ref{eq:rotor-basis-definition}) are
exactly the spatial rotations of nematic mean-field states, characterized by
the order parameter in Eq.~(\ref{eq:degenerate-nematic-states}). More
precisely, each distinct mean-field state with an order parameter of the form
$R(g) \chi_\text{n}(\eta)$ (where $g \in \text{SO}(3)$ ranges over all
rotations, $R(g)$ is the matrix corresponding to $g$ in the 5-dimensional
representation, and $\chi_\text{n}(\eta)$ are the 5-component order parameters
of Eq.~(\ref{eq:degenerate-nematic-states})), can be expressed in the form of
Eq.~(\ref{eq:rotor-basis-definition}) with $\vec{\Omega}$ belonging to exactly
one pair of diametrically opposite points on the 4-sphere.

Due to the overcompleteness of the basis, any state $\ket{\Psi}$ can be
expressed as $\ket{\Psi} = \int_{\mathcal{S}^4} \dd^4\Omega\,\psi(\vec{\Omega})
\ket{\vec{\Omega}}$.  For any particle-conserving Hamiltonian $\hat{H}$ it
is also possible to find an operator $\mathcal{H}$, interpreted as a new
Hamiltonian acting on $\psi(\vec{\Omega})$, such that $\hat{H} \ket{\Psi} =
\int_{\mathcal{S}^4} \dd^4\Omega\, \left( \mathcal{H} \psi
\right)(\vec{\Omega})\ket{\vec{\Omega}}$ is exact. The mapping consists of
letting
\begin{equation}
	\hat{b}_i^\dagger \hat{b}_j \rightarrow \left( N + 5
	\right)\Omega_i \Omega_j - \Omega_j \nabla_i - \delta_{i j}
	\label{eq:rotor-transformation-rule}
\end{equation}
as has been derived in \cite{Barnett2010}.  The operator
$\vec{\nabla}$ acts as the gradient on functions defined on the 4-sphere and
yields a vector field, lying in the tangent space to the sphere at each point.
If we think of a function $f$ on $\mathcal{S}^4$ as a restriction of a function
on a broader subset of $\mathbb{R}^5$, the action can be expressed as
$\vec{\nabla}f = \frac{\partial f}{\partial \vec{\Omega}} - \left( \vec{\Omega}
\cdot \frac{\partial f}{\partial \vec{\Omega}} \right) \vec{\Omega}$.

Using the rule in Eq.~(\ref{eq:rotor-transformation-rule}), the norm property
$\vec{\Omega}\cdot \vec{\Omega} = 1$ and the intuitive identity
$\vec{\Omega}\cdot\vec{\nabla} = 0$, we find that $\vec{\hat{b}}^\dagger S_{i
j}^{(-)} \vec{\hat{b}} \rightarrow \vec{\Omega}^T S_{i j}^{(-)} \vec{\nabla} =
-\ii L_{i j}$ where $L_{i j} = -\ii \left( \Omega_i \nabla_j - \Omega_j
\nabla_i \right)$ is the 5-dimensional generalization of angular momentum. This
in turn implies $\vec{\hat{b}}^\dagger M^i \vec{\hat{b}} \rightarrow
\vec{\Omega}^T  M^i \vec{\nabla}$. We also find  $\hat{A}^\dagger \hat{A}
\rightarrow \nabla^2 + (N^2 + 3 N)$.  Putting this all together and dropping
constant terms yields
\begin{eqnarray}
	\mathcal{H} &=& \frac{g_2}{2 N}\nabla^2 + \frac{g_1}{2 N}
	\Omega_\alpha M^i_{\alpha \beta} \nabla_\beta\,
	\Omega_\gamma M^i_{\gamma \delta} \nabla_\delta\nonumber\\
	{}&+&q\left( N+5 \right) \vec{\Omega}^T Q\, \vec{\Omega} -
	q\, \vec{\Omega}^T Q \vec{\nabla}
		\label{eq:rotor-nonHermitian-hamiltonian}
\end{eqnarray}
Recall that $Q \equiv (M^z)^2$. To discuss individual parts of the Hamiltonian
we will also refer to the operator multiplying $\frac{g_1}{2 N}$ as
$\vec{M}^2$.

When $q=0$, the resulting Hamiltonian is Hermitian, with the ground state
uniformly delocalized about the 4-sphere, which corresponds, loosely speaking,
to a condensate of singlet pairs in accord with previous results
\cite{Koashi2000,Ueda2002}.  It is interesting to comment on this result in
light of the recent publication by Jen and Yip \cite{Jen2015} who pointed out
that even though na\"ive averaging of nematic states over rotations in all of
SO$(3)$ produces the correct groundstates for confined antiferromagnetic spin-1
bosons, extending this to spin-2 does not work, as the singlet is no longer
unique in this case.  The rotor mapping demonstrates that the correct state can
in fact be obtained by averaging over the associated 4-sphere.

In the general case, the obtained Hamiltonian is not Hermitian. When $g_1 = 0$
or when $ N |q| \gg |g_{1,2}|$ a   similarity transform may be enacted which
renders the transformed Hamiltonian Hermitian and which depends only on the
position operators $\Omega_i$.  This is the topic of the next sections. A
Hermitianizing similarity transform has also been identified for the general
case, but it is rather different from the one considered in this article, as it
is a complicated function of the Laplacian operator. It is not at present clear
whether that approach leads to similar calculational simplifications as
obtained in this article and its investigation is deferred to a future
publication.

For completeness, and since we shall not be using the result further, we derive
in the following brief subsection the form of the Hermitianized Hamiltonian for
the special case $g_1 = 0$.

\subsection{Hermitianizing transform at $g_1 = 0$}
\label{sec:special-case-hermitianizing}

In this special case the Hamiltonian $\mathcal{H}$ of
Eq.~(\ref{eq:rotor-nonHermitian-hamiltonian}) simplifies considerably as
$\vec{M}^2$, arguably the most complicated term, is not present. We assume the
correct similarity transform is of the form $\ee^{S}$ where $S =
S(\vec{\Omega})$ is a function of only the position operators. We seek $S$ such
that $\mathcal{H}^H_0 \equiv \ee^{-S} \mathcal{H} \ee^S$ is Hermitian. The
$\vec{\Omega}^T Q \vec{\Omega}$ term
Eq.~(\ref{eq:rotor-nonHermitian-hamiltonian}) is invariant under this
transformation. The Laplacian transforms as $\ee^{-S} \nabla^2 \ee^S = \nabla^2
+ (\nabla^2 S) + \left| \vec{\nabla} S \right|^2 + 2(\vec{\nabla} S)^T
\vec{\nabla}$ and the final non-Hermitian term of
Eq.~(\ref{eq:rotor-nonHermitian-hamiltonian}) picks up a Hermitian $-q
\vec{\Omega}^T Q (\vec{\nabla} S)$ term. Gathering the evidently non-Hermitian
terms and demanding that their sum be zero yields the condition
\begin{equation}
	\left( \frac{g_2}{N}\left( \vec{\nabla} S \right)^T - q \vec{\Omega}^T
	Q \right) \vec{\nabla} = 0.
	\label{eq:hermitian-condition}
\end{equation}
While one could make progress by formally solving a differential equation for
$S$ on the 4-sphere derived from the above, we avoid the tedious aspects of
doing so by positing that $S = \vec{\Omega}^T X \vec{\Omega}$ for some matrix
$X$. Inserting the ansatz into condition~(\ref{eq:hermitian-condition}) and
recalling that $\vec{\Omega}\cdot \vec{\nabla} = 0$, we see that $X = \frac{q
N}{2 g_2} Q$ indeed satisfies the condition. By defining $\rho_a \equiv
\sqrt{\Omega_2^2 + \Omega_3^2}$ and $\rho_b \equiv \sqrt{\Omega_4^2 +
\Omega_5^2}$ this may be put into simple terms as $S = \frac{q N}{2
g_2}(\rho_a^2 + 4 \rho_b^2)$. After expanding out $S$ in the remaining terms
added by the transformation, the final Hamiltonian is found to be
\begin{eqnarray}
	&&\mathcal{H}^H_0 = \frac{g_2}{2 N} \nabla^2 + q \left( N + \frac{5}{2}
	- \frac{q N}{2 g_2} \right) \rho_a^2\nonumber\\
	&&{} + 4 q \left( N + \frac{5}{2} -
	\frac{2 q N}{g_2} \right) \rho_b^2 + \frac{q^2 N}{2 g_2} \left(
	\rho_a^2 + 4 \rho_b^2 \right)^2
	\label{eq:g1-zero-hermitian-hamiltonian}
\end{eqnarray}

\section{Large \large$ N |q|$\small\ limits}
\label{sec:large-q-limits}
\subsection{Large positive $ N q$ regime}
\label{sec:large-positive-q}

For large positive $N q$ the dominant $q \hat{Z} = q \hat{a}_\alpha^\dagger
(F^z)^2_{\alpha \beta} \hat{a}_\beta$ term in Hamiltonian
(\ref{eq:many-body-hamiltonian}) is minimized for the state $a_0^{\dagger N}
\ket{0} = b_1^{\dagger N} \ket{0} = \frac{1}{2}
\int_{\mathcal{S}^4}\dd^4\Omega\,\prod_{i=2}^{5} \delta(\Omega_i) \ket{\vec{\Omega}}$, suggesting 
\footnote{This state minimizes the $q \hat{Z}$ term among all states with $N$ particles, without regard to fixing
$F_z$. The state obviously has $F_z = 0$, as does the limiting
state~(\ref{eq:large-negative-q-mean-field-state}) in the negative $q$ regime,
suggesting the $\hat{F}_z$ nullspace as a particularly natural choice. 
We consider $F_z$ to be fixed at 0 for the remainder of
Sec.~\ref{sec:large-q-limits}.} 
that the low-lying exact eigenstates are
tightly localized about the $\Omega_1 = \pm 1$ poles. As shown in
\cite{Barnett2011} the wave function has to have parity $\left( -1 \right)^N$,
so we may restrict our attention to the region about one of the poles and infer
the wave function's behaviour about the other by symmetry. We choose to expand
about the $\Omega_1 = +1$ pole, motivating the reparameterization
\begin{equation}
	\vec{\Omega} = ( \sqrt{1 - \vec{x}^2}, \vec{x})^T.
	\label{eq:large-positive-q-coords}
\end{equation}
We take the indices of $\vec{x}$ to run from $2$ to $5$ to avoid excessive
arithmetic in subscripts. Next assume that low-lying states are of the form
\begin{equation}
	\psi_n(\vec{x}) = h_n(\vec{x}) \ee^{-\frac{N}{2}
	\vec{x}^T \Gamma \vec{x} }
	\label{eq:large-positive-q-wavefunction-form}
\end{equation}
where $n$ is a generic (multi)index label, $\Gamma = \text{diag}\left(
\gamma_2, ..., \gamma_5 \right)$ is some diagonal matrix and $h_n$ are some
residual functions of sub-exponential growth such as Hermite polynomials. The
overall factor of $N$ was extracted for later convenience. The diagonal
elements of $N \Gamma$ can be interpreted as inverse squared oscillator lengths
$\xi_{i 0}$ for the $i$-th direction, i.e.~$N \gamma_i = \xi_{i 0}^{-2}$. Our
assumption of tight localization amounts to the condition $\xi_{i 0} \ll 1$,
which  has to be checked for consistency at the end of the calculation.  Since
$\left\langle x_i^n \partial_j^m \right\rangle \lesssim \xi_{i 0}^n /
\xi_{j 0}^m$ \footnote{Due to wave function parity such expectation values may be
much less or even vanish, but the stated quantity is the upper limit on their order of
magnitude.}, this allows us to simplify the
Hamiltonian~(\ref{eq:rotor-nonHermitian-hamiltonian}) by keeping only the
lowest $\xi_{i 0}$ terms multiplied by each of $\frac{g_{1,2}}{N}$, $q$, and
$N q$.

The goal now is to express
Hamiltonian~(\ref{eq:rotor-nonHermitian-hamiltonian}) in terms of $x_i$
and $\partial_i \equiv \frac{\partial}{\partial x_i}$. The former follows
from the coordinate definitions in Eq.~(\ref{eq:large-positive-q-coords}) while
the latter follows from computing $\nabla_\alpha=\hat{{\bf x}}_\alpha \cdot \nabla$
where $\hat{{\bf x}}_\alpha$ is a unit vector and $\nabla$ is the gradient operator expressed
in terms of the new coordinate system.
This leads to
\begin{eqnarray}
	\nabla_1 &=& -\sqrt{1-\vec{x}^2}\;\vec{x}\cdot
	\vec{\partial}\nonumber\\
	\nabla_i &=& \partial_i-x_i\, \vec{x}\cdot
	\vec{\partial}\qquad\text{for}\;i>1
	\label{eq:large-positive-q-derivatives}
\end{eqnarray}
Carrying out the necessary index algebra and truncating at the lowest
order $\xi_{i 0}$ terms yields the simple expressions
\begin{align}
	\nabla^2 &\simeq \vec{\partial} \cdot \vec{\partial} & \vec{\Omega}^T
	Q\, \vec{\Omega} &\simeq \vec{x}^T Q' \vec{x}\nonumber\\
	\vec{M}^2 &\simeq -3 \left( \partial_2^2 + \partial_3^2 \right) & \vec{\Omega}^T Q
	\vec{\nabla} &\simeq \vec{x}^T Q' \vec{\partial}
	\label{eq:simple-large-q-building-blocks}
\end{align}
where $Q' \equiv \text{diag}\,(1,1,4,4)$ is $Q$ with the first row and column
omitted. Putting this all together and letting $p_i = -\ii\, \partial_i$, we
obtain an approximate Hamiltonian $\mathcal{H}_+ = \sum_{i=2}^5 \mathcal{H}_i$
where
\begin{equation}
	\mathcal{H}_i = \frac{A_i}{2 N}p_i^2 + \frac{N B_i}{2} x_i^2 -
	\ii C_i x_i p_i
	\label{eq:large-q-single-direction-hamiltonian-form}
\end{equation}
where we do not sum over any repeated indices. The various constants in this
Hamiltonian are as follows
\begin{alignat}{2}
	A_{2,3} &= 3 g_1 + |g_2| \quad & A_{4, 5} &= |g_2|\nonumber\\
	C_{2,3} &= q & C_{4,5} &= 4 q\nonumber\\
	B_i &= t_N C_i & \nu_i &\equiv C_i/A_i\nonumber\\
	t_N &\equiv 2 + 5/N
	\label{eq:large-positive-q-pretransf-constants}
\end{alignat}
where $t_N$ and $\nu_i$ are  introduced for the purpose of later notation. This
allows us to treat each direction individually.  Following reasoning analogous
to that of Sec.~\ref{sec:special-case-hermitianizing} and applying the
similarity transform $\mathcal{H}^H_+ = \ee^{-S} \mathcal{H}_+ \ee^S$ with $S =
- N \sum_i \nu_i x_i^2/2$ we obtain a Hermitian sum of four independent
harmonic oscillator Hamiltonians, i.e., a Hamiltonian of the same form as
Eq.~(\ref{eq:large-q-single-direction-hamiltonian-form}) but with new constants
$A_i'=A_i$, $B_i'=B_i+C_i^2/A_i$ and $C_i'=0$.

This allows us to simply read off mode energies and oscillator lengths. They
are
\begin{eqnarray}
	\Delta E_i &=& C_i \sqrt{1 + t_N/\nu_i}\nonumber\\
	\xi_i^{-2} &=& N \nu_i \sqrt{1 + t_N / \nu_i}\nonumber\\
	N \gamma_i = \xi_{i 0}^{-2} &=& N \nu_i \left( 1+ \sqrt{1 + t_N/\nu_i} \right)
	\label{eq:large-q-energy-osc-lengths}
\end{eqnarray}
where $\xi_i$ are the oscillator lengths of the Hermitianized Hamiltonian
whereas $\xi_{i 0}$ are those of the original non-Hermitian Hamiltonian. The
solutions are indeed of the form assumed in
Eq.~(\ref{eq:large-positive-q-wavefunction-form}). Referring to
Eq.~(\ref{eq:large-positive-q-pretransf-constants}) allows us to verify that
$\xi_{i 0} \ll 1$ and the consistency of our approach when $N q \gg \left|
g_{1,2} \right|$.

The obtained mode energies agree very well with the numerically obtained
spectrum. As an illustration, the largest relative discrepancy among the $100$
lowest analytically and numerically obtained energies at $N=100,g_1 = |g_2|, q
= 100 |g_2|$ is $1.1$ percent.  The accuracy of the oscillator lengths, and the
wave function in general, is discussed in Sec.~\ref{sec:q-overlaps}.

It is interesting to note that the four modes agree exactly with the continuum
Bogoliubov mode energies at zero momentum, minus the density
mode.~\cite{Stamper-Kurn2013} We believe this to be a nontrivial result as the
number of particles $N$ does not neccessarily have to be large.  Nevertheless,
the limiting state about which we are expanding is of the mean-field form. 

The rotor framework is also capable of describing excitations about fragmented
states.  This will be demonstrated   in the following subsection.  As stated
previously, such excitations are outside the reach of conventional Bogoliubov
analysis.

\subsection{Large negative $N q$ regime}
\label{sec:large-negative-q}

For large negative values of $N q$, i.e., when $-N q \gg \left| g_{1,2} \right|$,
the dominant $q \hat{Z}$ term in Hamiltonian~(\ref{eq:many-body-hamiltonian})
is minimized for the state
\begin{eqnarray}
\left( a_2^\dagger  a_{-2}^\dagger \right)^{N/2} \ket{0} 
	&\propto& \int\dd\varphi\,\left( \ee^{\ii \varphi}a_2^\dagger +
		\ee^{-\ii \varphi} a_{-2}^\dagger \right)^N \ket{0} 
			\label{eq:large-negative-q-mean-field-state}
		\\
	&\propto& \int \dd\varphi\,\left( \cos\varphi\,b_4^\dagger +
		\sin\varphi\,b_5^\dagger \right)^N \ket{0}\nonumber\\
		&\propto& \int \dd^4 \Omega \,\delta\left( \Omega_1
		\right) \delta\left( \Omega_2 \right) \delta\left( \Omega_3
		\right) \ket{\vec{\Omega}}
		\nonumber
\end{eqnarray}
Note that line one of the above equation clearly demonstrates that we are
working with a fragmented state, with two macroscopically occupied
single-particle states for large $N$. As mentioned before, the rotor
mapping is of particular utility here.

An appropriate reparameterization in this case is
\begin{eqnarray}
	\Omega_i &=& x_i\quad\qquad\;\text{for }i = 1, 2, 3\nonumber\\
	\left( \Omega_4, \Omega_5 \right) &=& \sqrt{1 - \vec{x}^2} \left(
	\cos \varphi, \sin \varphi \right)\quad
	\label{eq:large-negative-q-coordinates-with-r}
\end{eqnarray}
where we have reused the label $x$ from the $N q \gg \left| g_{1,2} \right|$
case for three of the coordinates and introduced the angular variable $\varphi$
as the fourth. Further reusing notation from the previous subsection, we
assume low-energy states can be written as
\begin{equation}
	\psi_n(\vec{x},\varphi) = h_n(\vec{x},\varphi)
	\ee^{-\frac{N}{2} \vec{x}^T \Gamma \vec{x} }
	\label{eq:large-negative-q-wavefunction-form}
\end{equation}
in analogy with Eq.~(\ref{eq:large-positive-q-wavefunction-form}) for large
positive $N q$. Here $h_n$ is of subexponential growth in $|\vec{x}|$ and
periodic in $\varphi$ and $\Gamma = \text{diag}\,(\gamma_1, \gamma_2,
\gamma_3)$. We again assume the $\xi_{i 0} \equiv (N \gamma_i)^{-1/2}$ are
small, allowing us to keep only the lowest $\xi_{i 0}$ terms multiplied by each
of $\frac{g_{1,2}}{N}$, $q$ and $N q$.  Additionally, we assume that the wave
function is not  localized  in the $\varphi$ direction, so that
$\partial_\varphi \equiv \frac{\partial}{\partial \varphi}$ is of order 1,
in the sense that its matrix elements with low-lying states are at most of
order 1.

Again let $\partial_i \equiv \frac{\partial}{\partial x_i}$ and define
$\vec{\partial} \equiv \left( \partial_1, \partial_2, \partial_3 \right)$,
where we note that $\vec{\partial}$ does not contain $\partial_\varphi$. The
gradient components are found to be
\begin{eqnarray}
	\nabla_i &=& \partial_i - x_i \, \vec{x} \cdot \vec{\partial}
	\quad \text{for }i = 1, 2, 3\nonumber\\
	\nabla_4 &=& -\sqrt{1 - \vec{x}^2} \cos\varphi \, \vec{x}
	\cdot \vec{\partial} - \frac{\sin\varphi}{\sqrt{1 - \vec{x}^2}}
	\partial_\varphi\nonumber\\
	\nabla_5 &=& -\sqrt{1 - \vec{x}^2} \sin\varphi \, \vec{x}
	\cdot \vec{\partial} + \frac{\cos\varphi}{\sqrt{1 - \vec{x}^2}}
	\partial_\varphi
	\label{eq:large-negative-q-derivatives}
\end{eqnarray}
Expressing components of Hamiltonian~(\ref{eq:rotor-nonHermitian-hamiltonian})
in terms of $\vec{x}, \varphi$ and their partial derivatives and
truncating higher order $\xi_{i 0}$ terms yields
\begin{align}
	\nabla^2 &\simeq \vec{\partial} \cdot \vec{\partial} &
	\vec{\Omega}^T Q\, \vec{\Omega} &\simeq - \vec{x}^T Q'' \vec{x}\nonumber\\
	\vec{M}^2 &\simeq -\partial_2^2 - \partial_3^2  &
	\vec{\Omega}^T Q \vec{\nabla} &\simeq - \vec{x}^T Q'' \vec{\partial}
	\label{eq:simple-negative-q-building-blocks}
\end{align}
where $Q'' =\text{diag}\, (4, 3, 3)$ is $(4\,\mathbb{1} - Q)$ with the last two
columns and rows omitted. This leads to $\mathcal{H}_- = \sum_{i=1}^3
\mathcal{H}_i$ with $\mathcal{H}_i$ of the same form as in
Eq.~(\ref{eq:large-q-single-direction-hamiltonian-form}) and the relevant
constants defined as:
\begin{alignat}{2}
	A_1 &= |g_2| & A_{2, 3} &= (g_1 + |g_2|)\nonumber\\
	C_1 &= 4 |q| & C_{2, 3} &= 3 |q|\nonumber\\
	B_i &= t_N C_i \quad & \nu_i &\equiv C_i/A_i
	\label{eq:large-negative-q-pretransf-constants}
\end{alignat}
with $t_N$ as in Eq.~(\ref{eq:large-positive-q-pretransf-constants}). The rest
of the calculation proceeds as in the previous section, again leading to
Eq.~(\ref{eq:large-q-energy-osc-lengths}) for $i = 1,2,3$, evaluated with the
above constants, and a validation of our assumptions of localized states.

Again, the mode energies are in excellent agreement with the numerics, with the
largest relative discrepancy among the first $100$ lowest energies at $N = 100,
g_1 = |g_2|, q = -100 |g_2|$ equal to $0.16$ percent.

\subsection{Wave function overlaps}
\label{sec:q-overlaps}

Besides facilitating the analytical derivation of excitation energies, the
rotor mapping also yields insightful information on the wave functions
themselves.  The associated 4-sphere often provides a more intuitive picture of
the wave function than the original second-quantized operator picture.

In this section we investigate the overlap of the ground state wave functions
with arbitrary values of $q$  with wave functions in the limit of large
$N|q|$.  The ground state wave functions will be computed in two ways.
In the first approach, we use the rotor mapping while with the second approach
we use exact diagonalization for modest numbers of total particles.  We label
the wave functions with the limiting large-$N|q|$ values as
\begin{align*}
\ket{\psi^\infty_+} &= \frac{1}{\sqrt{N!}}  (\hat{a}_0^\dagger)^N \ket{0} \\ 
\ket{\psi^\infty_-} &= \frac{1}{(N/2)!}  (\hat{a}_2^\dagger \hat{a}_{-2}^\dagger)^N \ket{0} 
\end{align*}
which are appropriate for large positive and large negative $Nq$, respectively.
The first state has a clear correspondence to the mean-field uniaxial nematic
state oriented along the $z$-axis (c.f.\ Fig.\  \ref{fig:mf-phase-diagram}).
The second fragmented state can be viewed as an equal-weight superposition of
all square biaxial nematic states lying in the $xy$ plane, as is evident from
Eq.~(\ref{eq:large-negative-q-mean-field-state}).  One may also view
$\ket{\psi^\infty_-}$ as the $F_z = 0$ component of any of these mean-field
ground states.  For large positive  or negative $Nq$, one expects a large
overlap of the ground state with $\ket{\psi_+^\infty}$ or
$\ket{\psi_-^\infty}$, respectively. On the other hand, for moderate $Nq$, one
may ask if any relic of the order-by-disorder phenomenon present in the
continuum case, as shown in Fig.~\ref{fig:mf-phase-diagram}, remains.

The simplest expressions for the overlaps may be obtained in the regime where
$N \gg 1$ and $|q|$ is not much smaller than either $|g_1|$ or $|g_2|$.  We
restrict our attention to this case in the following.  This is slightly more
restrictive than the condition of the previous section, namely  $N q \gg
|g_{1,2}|$.  For the case when $N q \gg |g_{1,2}|$, but $N$ is not large
compared to unity, the analysis is complicated by the interplay between
asymptotic series convergence and the applicability of extending Gaussian
integration limits to infinity.

We define the overlap of two possibly unnormalized states $\ket{a}$ and $\ket{b}$
as $\left( a \middle| b \right) \equiv \left| \braket{a}{b}
\right|/\sqrt{\braket{a}{a}\braket{b}{b}}$. States are completely determined by
their wave function in the overcomplete basis and we follow the convention of
labelling states of the original Hamiltonian by the same label as their rotor
wave functions. That is \begin{equation} \ket{\psi} \equiv \int_{\mathcal{S}^4}
\dd\vec{\Omega}\, \psi(\vec{\Omega}) \ket{\vec{\Omega}}.
\label{eq:rotor-to-original-labelling-convention} \end{equation} We label the
ground states as obtained through the rotor mapping by
$\ket{\psi_\pm^\text{R}}$. The sign in the subscript indicates whether we
expanded Hamiltonian~(\ref{eq:rotor-nonHermitian-hamiltonian}) about the
large positive- or large negative-$N q$ limiting state. We label the numerically
obtained ground states by $\ket{\psi^\text{N}}$.

While the overcompleteness of the basis did not manifest itself significantly
in calculating the spectrum, it does affect calculations involving the
eigenfunctions. As is simple to verify from the definition of
$\ket{\vec{\Omega}}$ states, $\braket{\vec{\Omega}_1}{\vec{\Omega}_2} = \left(
\vec{\Omega_1} \cdot \vec{\Omega_2} \right)^N$.  In the thermodynamic limit,
one can express this inner product in terms of delta functions on the
four-sphere.  However, for finite $N$, overlaps must be computed by means of
double integrals over the 4-sphere:
\begin{equation}
	\braket{\psi_a}{\psi_b} = \int_{\mathcal{S}^4}
	\dd\vec{\Omega}_1\,\int_{\mathcal{S}^4}\dd\vec{\Omega}_2\,
	\psi_a^*(\vec{\Omega}_1)\psi_b(\vec{\Omega}_2) \left( \vec{\Omega}_1
	\cdot \vec{\Omega}_2 \right)^N
	\label{eq:double-integral}
\end{equation}

\begin{figure}
	\includegraphics[width=\columnwidth]{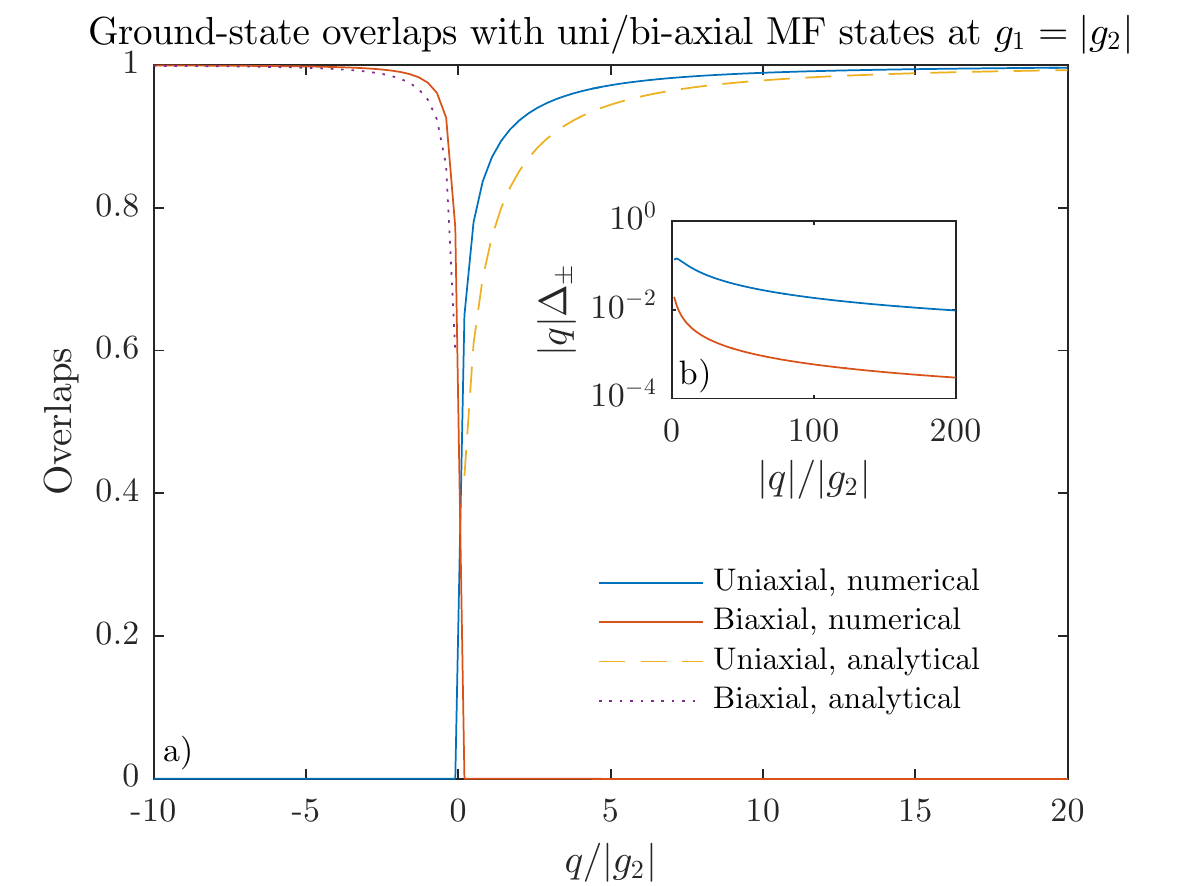}
	\caption
	{a) 
	Comparison of numerically and analytically obtained overlaps,
	$\left( \psi_\pm^\infty \middle| \psi^{\text{N}} \right)$  and
	$\left( \psi_\pm^\infty \middle| \psi_\pm^{\text{R}} \right)$,
	represented by solid and dashed lines in the plot, respectively, where
	$\psi^{\infty}_{\pm}$ denote the limiting states. 
	For large $|q|/|g_2|$ both
	tend to one or zero.  Inset b) demonstrates that $|q|\Delta_\pm$, where
	$\Delta_\pm$ is defined under Eq.~(\ref{eq:positive-q-overlaps-final}),
	tends to zero with increasing $|q|$, implying that our analytical and
	numerical expressions agree to at least first order in asymptotic
	expansion.}
	\label{fig:mean-field-rotor-quad-Zeeman-overlaps}
\end{figure}
\subsubsection{The case of positive $q$}

Here we reuse the $\vec{x}$ coordinates of Sec.~\ref{sec:large-positive-q} as
defined in Eq.~(\ref{eq:large-positive-q-coords}). We  integrate over only half
of the 4-sphere, as this is less cluttered by trivial (anti)symmetrizations.
The relevant wave functions in the rotor picture
(\ref{eq:rotor-to-original-labelling-convention}) are $\psi_+^\infty(\vec{x}) =
\delta^{(4)}(\vec{x})$ and $\psi_+^\text{R}(\vec{x})$. The latter is of the
form of Eq.~(\ref{eq:large-positive-q-wavefunction-form}) with $h_n$ equal to
1, i.e., $\psi_+^\text{R}(\vec{x}) = \exp \left( -\frac{N}{2} \vec{x}^T \Gamma
\vec{x} \right)$, with $\Gamma = \text{diag}\, (\gamma_2, \gamma_3, \gamma_4,
\gamma_5)$ and $\gamma_i$ as expressed in
Eq.~(\ref{eq:large-q-energy-osc-lengths}), evaluated with the values given by
Eq.~(\ref{eq:large-positive-q-pretransf-constants}).

In the new coordinates we have $\dd\vec{\Omega} = \dd\vec{x}/\sqrt{1 -
\vec{x}^2}$ and the dot product between vectors on the four sphere is expressed
as $\vec{\Omega}_1 \cdot \vec{\Omega}_2 = \sqrt{\left( 1 - \vec{x}_1^2 \right)
\left( 1 - \vec{x}_2^2 \right) + \vec{x}_1 \cdot \vec{x}_2}$.  Assuming tight
localization about $\vec{x} = 0$, the main contribution to the integral will
come from that region and we may extend the boundary of integration from
$|\vec{x}| = 1$ to $|\vec{x}| \rightarrow \infty$.  The denominator of the new
integration measure varies relatively slowly, so we may set it to its
value at $\vec{x} = 0$.

Due to the simplicity of $\psi_+^\infty$, a straightforward calculation gives
\begin{eqnarray}
	\braket{\psi_+^\infty}{\psi_+^\infty} &=& 1\\
	\braket{\psi_+^\infty}{\psi_+^\text{R}} &=& \int_{\mathbb{R}^4}
	\dd\vec{x}\, \left( 1 - \vec{x}^2 \right)^{\frac{N}{2}}
	\ee^{-\frac{N}{2}\vec{x}^T \Gamma \vec{x}}\nonumber\\
	&\simeq& \int_{\mathbb{R}^4} \dd \vec{x}\, \ee^{-\frac{N}{2}
	\vec{x}^T\left( \Gamma + \mathbb{1} \right) \vec{x}}\nonumber\\
	&=& \prod_{i=2}^5 \sqrt{\frac{2 \pi}{N \left( \gamma_i + 1 \right)}}.
	\label{eq:overlap-with-mean-field-simple-positive-q}
\end{eqnarray}
On the third line we approximated $1 - \vec{x}^2 \simeq \ee^{-\vec{x}^2}$,
permissible on account of tight localization.

Evaluation of $\braket{\psi_+^\text{R}}{\psi_+^\text{R}}$ involves the
approximation (valid due to the localized wave functions) ${\bf \Omega}_1 \cdot
{\bf \Omega_2} = \sqrt{\left( 1 - \vec{x}_1^2 \right) \left( 1 - \vec{x}_2^2
\right)} + \vec{x}_1 \cdot \vec{x}_2 \simeq 1 - \frac{\vec{x}_1^2}{2} -
\frac{\vec{x}_2^2}{2} + \vec{x}_1 \cdot \vec{x}_2 = 1 - \frac{1}{2}\left(
\vec{x}_1 - \vec{x}_2 \right)^2 = 1 - \vec{y}_2^2 \simeq \ee^{-\vec{y}_2^2}$
where we introduced new integration variables $\vec{y}_{1,2} \equiv \left(
\vec{x}_1 \pm \vec{x}_2 \right)/\sqrt{2}$. With these variables and the above
approximation, the integrand becomes $\exp \left[ -\frac{N}{2} \left(
\vec{y}_1^T \Gamma \vec{y}_1 + \vec{y}_2^T \left( \Gamma + 2 \mathbb{1} \right)
\vec{y}_2 \right) \right]$, leading to
$\braket{\psi_+^\text{R}}{\psi_+^\text{R}} = \prod_{i = 2}^5 \frac{2 \pi}{N}
\left[ \gamma_i \left( \gamma_i + 2 \right) \right]^{-1/2}.$

Combining the results of the previous paragraph and
Eq.~(\ref{eq:overlap-with-mean-field-simple-positive-q}), we find that the
total overlap $\left( \psi_+^\infty \middle| \psi_+^\text{R} \right)$ can be
expressed as a product of contributions from individual $x_i$-directions and
that the $i$-th direction contributes a factor of $\left[ \frac{\gamma_i\left(
\gamma_i + 2 \right)}{\left( \gamma_i + 1 \right)^2} \right]^{1/4}$. This
prompts us to define
\begin{equation}
	u_i^2 \equiv \frac{\left( \gamma_i + 1 \right)^2}{\gamma_i \left(
	\gamma_i + 2 \right)} = \frac{1}{2} \left( 1 + \frac{\nu_i +
	1}{\sqrt{\nu_i \left( \nu_i + 2 \right)}} \right)
	\label{eq:overlap-u-definition}
\end{equation}
where the rightmost expression was derived by expanding $\gamma_i$ in terms of
$\nu_i$ as in Eq.~(\ref{eq:large-q-energy-osc-lengths}) and letting $t_N \equiv
2 + 5/N \simeq 2$. The $\nu_i$ are defined in
Eq.~(\ref{eq:large-positive-q-pretransf-constants}) and are summarized here for
convenience:
\begin{equation}
	\nu_a \equiv \nu_{2, 3} = \frac{q}{3 g_1 + |g_2|} \qquad \nu_b \equiv
	\nu_{4, 5} = \frac{4 q}{|g_2|}.
	\label{eq:overlaps-positive-q-nus}
\end{equation}
Since each direction contributes a factor of $u_i^{-1/2}$, the total overlap is
\begin{equation}
	\left( \psi_+^\infty \middle| \psi_+^\text{R} \right) = u_a^{-1}
	u_b^{-1}.
	\label{eq:positive-q-overlaps-final}
\end{equation}

The overlap $\left( \psi_+^\infty\middle| \psi_+^\text{R} \right)$ is plotted
in the main panel of Fig.~\ref{fig:mean-field-rotor-quad-Zeeman-overlaps} for
$N=200$ particles and $g_1 = |g_2|$.  For comparison, we have used exact
diagonalization to determine the the wave function $\ket{\psi^N}$ and the
overlaps $(\psi^\infty_{\pm} | \psi^N)$ for the same parameter ranges.  As is
expected, for large positive $q$ both the analytical and numerical overlap
expressions approach unity for large $|q|$.  To show that the two agree in more
than just this obvious large-$q$ limit, we consider their asymptotic
expansions.  Let $f_\pm  =(\psi^\infty_{\pm} | \psi^R_{\pm}) =  1 +
\sum_{n=1}^{\infty}a_n q^{-n}$ and $g_\pm   =(\psi^\infty_{\pm} | \psi^N) =  1
+ \sum_{n=1}^{\infty}b_n q^{-n}$.  Define $\Delta_\pm \equiv \left| f_\pm -
g_\pm \right| = \left|\sum_{n=1}^\infty \left( a_n - b_n \right)
q^{-n}\right|$.  In the inset of
Fig.~\ref{fig:mean-field-rotor-quad-Zeeman-overlaps} we show that $q \Delta_+$
tends to zero with increasing $q$, implying that our analytical expressions
agree with the numerics to at least the first order in the asymptotic
expansion.

\subsubsection{The case of negative $q$}

For this subsection, we reuse the $\vec{x}$ and $\varphi$ coordinates of
Sec.~(\ref{sec:large-negative-q}) defined in
Eq.~(\ref{eq:large-negative-q-coordinates-with-r}). The limiting large and
negative $q$ rotor wave function is $\psi_-^\infty(\vec{x}, \varphi) =
\delta^{(3)}(\vec{x})$.  The finite-$q$ ground-state as obtained in
section~\ref{sec:large-negative-q} is $\psi_-^\text{R}(\vec{x}, \varphi) = \exp
\left( - \frac{N}{2} \vec{x}^T \Gamma \vec{x} \right)$, with the matrix $\Gamma
= \text{diag}\,(\gamma_1, \gamma_2, \gamma_3)$ as defined underneath
Eq.~(\ref{eq:large-negative-q-wavefunction-form}) and the $\gamma$ variables as
defined in Eq.~(\ref{eq:large-q-energy-osc-lengths}), evaluated at values from
Eq.~(\ref{eq:large-negative-q-pretransf-constants}).

In the new coordinates, one has $\dd\vec{\Omega} = \dd \varphi\, \dd \vec{x} /
\sqrt{1 - \vec{x}^2} \simeq \dd \varphi\, \dd \vec{x}$, with the last
approximation being permissible on account of localization, as in the positive
$q$ case. As before we may extend the $\vec{x}$ integration boundaries to
infinity. The range of integration in $\varphi$ is from $0$ to $2 \pi$.  The
dot product between vectors on the four sphere is  $\vec{\Omega}_1 \cdot
\vec{\Omega}_2 = \cos (\varphi_1 - \varphi_2) \sqrt{\left( 1 - \vec{x}_1^2
\right) \left( 1 - \vec{x}_2^2 \right)} + \vec{x}_1 \cdot \vec{x}_2$. Since the
considered wave functions do not depend on the $\varphi$ coordinate, we may
simplify integration over $\varphi_{1,2}$ by a change of variables. Defining
$\varphi \equiv \varphi_1 - \varphi_2$ and, say, $\varphi_2' \equiv \varphi_2$
allows us to immediately perform the now trivial $\varphi_2'$ integral to
obtain
\begin{eqnarray}
	&&\braket{\psi_a}{\psi_b} = 2 \pi \int_0^{2 \pi} \dd \varphi\,
	\iint_{\mathbb{R}^3} \dd \vec{x}_1 \, \dd \vec{x}_2 \, \psi_a^*
	(\vec{x}_1) \psi_b (\vec{x}_2) \times \nonumber\\
	&&\qquad \qquad\left( \cos \varphi \sqrt{\left( 1 - \vec{x}_1^2 \right)
	\left( 1 - \vec{x}_2^2 \right)} + \vec{x}_1 \cdot
	\vec{x}_2 \right)^N
	\label{eq:overlap-negative-q-general}
\end{eqnarray}
where $\psi_{a,b}$ are any wave functions that do not depend on the $\varphi$
variable, such as $\psi_-^\infty$ or $\psi_-^\text{R}$. Using this
expression and approximations analogous to those of
Eq.~(\ref{eq:overlap-with-mean-field-simple-positive-q}) the simpler integrals
are found to be:
\begin{eqnarray}
	\braket{\psi_-^\infty}{\psi_-^\infty} &=& 2 \pi \int_0^{2 \pi}
	\dd \varphi\, \cos^N \varphi \equiv \mathcal{N}_-\\
	\braket{\psi_-^\infty}{\psi_-^\text{R}} &\simeq& \mathcal{N}_-
	\int_{\mathbb{R}^4} \dd \vec{x}\, \ee^{-\frac{N}{2} \vec{x}^T\left(
	\Gamma + \mathbb{1} \right) \vec{x}}\nonumber\\
	&=& \mathcal{N}_- \prod_{i=1}^3 \sqrt{\frac{2 \pi}{N \left( \gamma_i +
	1 \right)}}.
	\label{eq:overlap-with-mean-field-simple-negative-q}
\end{eqnarray}

To calculate $\braket{\psi_-^\text{R}}{\psi_-^\text{R}}$, consider again the
factor $f \equiv \left( \cos \varphi \sqrt{\left( 1 - \vec{x}_1^2 \right)
\left( 1 - \vec{x}_2^2 \right)} + \vec{x}_1 \cdot \vec{x}_2 \right)^N$ of
Eq.~(\ref{eq:overlap-negative-q-general}). Due to the large exponent $N$, the
significant contributions to the integral will come from regions of maximum
$|\cos \varphi|$, that is for $\varphi \sim 0$ or $\pi$. In both regions, we
may expand $\cos \varphi$ to quadratic order and extend integration boundaries
to infinity, yielding a Gaussian integral in $\delta \varphi \equiv \varphi -
\varphi_0$ where $\varphi_0 = 0$ or $\pi$.  Also expanding the square roots and
keeping lowest order terms in $\vec{x}_{1, 2}$ and $\varphi$ yields
\begin{equation}
	f \simeq \exp \left[ -\frac{N}{2} \left( \delta\varphi^2 + 2
	\vec{y}_r^2 + \sum_{i=1}^2 \vec{y}_i^T \Gamma \vec{y}_i^T \right)
	\right]
	\label{eq:overlaps-negative-q-third-integral-intermediate}
\end{equation}
where $\vec{y}_{1, 2} \equiv \left( \vec{x}_1 \pm \vec{x}_2 \right)/\sqrt{2}$
as in the positive-$q$ case. The label $r$ equals 1 for the $\varphi_0 = \pi$
region and 2 for the $\varphi_0 = 0$ region. The integrals over $\vec{y}_{1,
2}$ are equal in both cases, and twice the $\delta\varphi$ integral is in fact
approximately equal to $\mathcal{N}_-$ of
Eq.~(\ref{eq:overlap-with-mean-field-simple-negative-q}), as can be verified by
applying the same approximate treatment of integration over $\varphi$ to
$\braket{\psi_-^\infty}{\psi_-^\infty}$. This leads to
$\braket{\psi_-^\text{R}}{\psi_-^\text{R}} = \mathcal{N}_- \prod_{i = 1}^3
\frac{2 \pi}{N} \left[ \gamma_i \left( \gamma_i + 2 \right) \right]^{-1/2}$.

Combining the above results and expressing everything in terms of $u_i$,
defined in Eq.~(\ref{eq:overlap-u-definition}) and evaluated at
\begin{equation}
	\nu_c \equiv \nu_1 = 4 \left|\frac{q}{g_2}\right| \qquad \nu_d \equiv
	\nu_{2, 3} = \frac{3 |q|}{g_1 + |g_2|},
	\label{eq:overlaps-negative-q-nus}
\end{equation}
summarized after Eq.~(\ref{eq:large-negative-q-pretransf-constants}),
ultimately yields
\begin{equation}
	\left( \psi_-^\infty \middle| \psi_-^\text{R} \right) =
	u_c^{-\frac{1}{2}} u_d^{-1}
	\label{eq:negative-q-overlaps-final}
\end{equation}
The main panel of Fig.~\ref{fig:mean-field-rotor-quad-Zeeman-overlaps} again
demonstrates that both numerical and analytical overlaps tend to 1 with
increasing $|q|$ while the inset shows that the convergence agrees to at least
the first order in the asymptotic expansion.

\section{Order-by-Disorder}
\label{sec:obd}

One of the most salient features seen in Fig.\
\ref{fig:mean-field-rotor-quad-Zeeman-overlaps} is the absence of the
order-by-disorder phenomenon which is present for the continuum case
\cite{Song2007,Turner2007}.  We note that while the analytical  expressions for
the overlaps $(\psi^\infty_{\pm} | \psi^R_{\pm})$ are  valid only for $|q|$
larger than either $|g_1|$ or $|g_2|$, the numerically computed overlaps
$(\psi^\infty_{\pm} | \psi^N)$ for modest particle number are valid for all
$q$. For the case of $g_1>0$, one might expect a tendency towards the uniaxial
nematic state for small $q$, but this is not exhibited in
Fig.~\ref{fig:mean-field-rotor-quad-Zeeman-overlaps}. Instead, for small $q$,
symmetry restoring fluctuations drive the system towards the singlet state
which is the true ground state for $q=0$ and finite particle number. Varying
$g_1/|g_2|$ only affects how quickly the ground state approaches the respective
limiting states. This effect is completely smooth in the whole nematic region:
the smaller $g_1/|g_2|$ is, the faster the ground states approach the limiting
mean-field states with increasing $|q|/|g_2|$, without any qualitative change
in behavior at $g_1 = 0$.

The lack of the order-by-disorder selection at the quantum / single-mode level
can be accounted for by the fact that the quadratic Zeeman potential breaks too
much symmetry.  Motivated by this, we consider an alternative external
potential.  Specifically, we consider a potential that replaces
\begin{align}
\label{newfield}
q \hat{Z} \rightarrow \lambda ( \hat{a}_1^\dagger \hat{a}_1 +
\hat{a}_{-1}^\dagger \hat{a}_{-1} )
\end{align}
in Hamiltonian~(\ref{eq:many-body-hamiltonian}).  Such a potential could be
realized with microwave fields.  We note  that within mean field theory, all
nematic states of the form (\ref{eq:degenerate-nematic-states}) are degenerate
under this external potential.  Considering the rotor mapping rule in
Eq.~(\ref{eq:rotor-transformation-rule}) one can see this propagates through
the mapping by changing the last line of
Hamiltonian~(\ref{eq:rotor-nonHermitian-hamiltonian}) to
\begin{equation}
	\mathcal{H}_\lambda = \lambda\left( \left( N + 5 \right)\left(
	\Omega_2^2 + \Omega_3^2 \right) - \Omega_2 \nabla_2 - \Omega_3 \nabla_3
	\right).
	\label{eq:lamba-hamiltonian-contrib}
\end{equation}

In the following we will perform an analysis on this model following the rotor
mapping of the previous Sections.  In the Appendix, an analysis of the
analogous continuum problem is discussed. 

\subsection{Rotor treatment}
\label{seq:obd-rotor}
The results of this section are similar to the large-$N |q|$ limit in that, for
sufficiently large $\lambda$ and depending on the sign of $g_1$, the rotor
wave function is localized either about the $\Omega_1$ pole or around the 4-5
equator of the 4-sphere. However, the localization width scales differently
with $N$ than in the quadratic Zeeman case, leading to important qualitative
differences. Localization at the pole (equator) also occurs at negative
(positive) $g_1$, which is in fact the opposite of the effect in the
continuum in the absence of an external potential.

For the calculations of this section we introduce a third, more general
coordinate system:
\begin{eqnarray}
	\vec{\Omega} = \left( 
	\begin{matrix}
		\cos\eta \sqrt{1 - \vec{x}^2}\\
		x_1 \cos\varphi - x_2 \sin\varphi\\
		x_1 \sin\varphi + x_2 \cos\varphi\\
		\sin\eta \cos 2\varphi \sqrt{1 - \vec{x}^2}\\
		\sin\eta \sin 2\varphi \sqrt{1 - \vec{x}^2}
	\end{matrix}
	\right)
	\label{eq:obd-general-coords}
\end{eqnarray}
This can be put into a more compact form  by using rotation matrices.  In
particular let  $R_{\alpha \beta}(\varphi)$ be the matrix  which rotates in the
$\alpha \beta$ plane by angle $\varphi$. Then the current coordinate system can
be written as $\vec{\Omega} = R_{2 3}(\varphi) R_{4 5}(2\varphi) R_{1 4}(\eta)
\left( \sqrt{1 - \vec{x}^2}, x_1, x_2, 0, 0 \right)^T$. Note that $R_{2
3}(\varphi) R_{4 5}(2\varphi) = \exp(-\ii \varphi M^z)$. Recalling that each
point of the 4-sphere is associated with a spatial rotation of a mean-field
nematic state, the $\eta$ coordinate is seen to correspond exactly to the
$\eta$ parameterizing the accidentally degenerate family of nematic states in
Eq.~(\ref{eq:degenerate-nematic-states}), while $\varphi$ and $\vec{x}$
determine their spatial orientations.

As usual we consider the $\hat{F}_z$-nullspace, meaning that our wave functions
will be independent of $\varphi$. By further observing factors of $N$ in
Hamiltonian~(\ref{eq:rotor-nonHermitian-hamiltonian}) expanded in
coordinates~(\ref{eq:obd-general-coords}) we can infer that low-lying wave
functions are again localized on the scale of order $N^{-1/2}$ in the $x$
variables. Assuming that $\eta$ is localized about some $\eta_0$ and denoting
$\delta\eta \equiv \eta - \eta_0$, we may also infer that low-lying states are
localized in $\delta\eta$ on a scale of order $N^{-1/4}$, subject to some
consistency criteria.  This allows us to separate the Hamiltonian into two
parts, $\hat{\mathcal{H}}_0$ of order $1$ and $\hat{\mathcal{H}}_\eta$ of orders between $N^{-1/4}$
to $N^{-3/4}$, and we discard terms of higher order in $1/N$. For compact
notation introduce matrices $A(\eta) \equiv \left( 1 + 2 g_1/|g_2| \right)
\mathbb{1} + B(\eta)$ with $B(\eta) = \frac{g_1}{|g_2|} \text{diag} \left( \cos
2\eta + \sqrt{3} \sin 2\eta, \cos 2\eta - \sqrt{3} \sin 2\eta \right)$. Denote
$\partial_i \equiv \frac{\partial}{\partial x_i}$ and $\partial_\eta \equiv
\frac{\partial}{\partial \eta} = \frac{\partial}{\partial \delta\eta}$. Let
$L_x \equiv -\ii\left( x_1 \partial_2 - x_2 \partial_1 \right)$ and $T_1 \equiv
x_1 \partial_1 + x_2^2 \partial_1^2 - (1 \leftrightarrow 2)$. Then we may write
\begin{eqnarray}
	\hat{\mathcal{H}}_0 &=& - \frac{|g_2|}{2 N} A_{i j}(\eta_0) \partial_i \partial_j +
	\lambda N \vec{x}^2 - \lambda \vec{x} \cdot \vec{\partial}\nonumber\\
	\hat{\mathcal{H}}_\eta &=& - \frac{|g_2|}{2 N} \left[ \partial_\eta^2 + \left(
	\cot \eta - B'_{i j}(\eta_0) x_i \partial_j \right)
	\partial_\eta\right.\nonumber\\
	&&{}+ \delta\eta B'_{i j}(\eta_0) \partial_i \partial_j + \frac{1}{2}
	\delta\eta^2 B''_{i j}(\eta_0) \partial_i \partial_j\nonumber\\
	&&\left.{}-\frac{\csc^2 \eta}{4} L_x^2 \right] + \frac{g_1}{2 N}
	\sqrt{3} \csc \eta\, T_1.
	\label{eq:zeroth-first-order-obd-hamiltonian}
\end{eqnarray}
The last line is of a non-negligible order only when the distance between
$\eta_0$ and $0$ or $\pi$ is of the order of $N^{-1/4}$ or less.

Noting that $\hat{\mathcal{H}}_0$ does not depend on $\delta\eta$, we may tackle the
above with degenerate perturbation theory.  First we note that $\hat{\mathcal{H}}_0$ may
be brought to Hermitian form by applying the similarity transformation
$\ee^{-S}\hat{\mathcal{H}}_0 \ee^S$ where
\begin{equation}
	S = - \frac{N \lambda}{2 |g_2|} \vec{x}^T A(\eta_0)^{-1} \vec{x}.
	\label{eq:obd-similarity-transform}
\end{equation}
The transformed Hamiltonian has the ground state energy
\begin{equation}
	E_0(\eta_0) = \lambda\left( 1 + \frac{1}{2} \text{Tr}\, \sqrt{\mathbb{1} +
	\frac{2|g_2|}{\lambda}A(\eta_0)} \right).
	\label{eq:obd-zeroth-order-gs-energy}
\end{equation}
and ground state eigenfunction 
\begin{align}
	\label{eq:obd-zeroth-order-x-wavefunction}
	\psi_0(\vec{x}) =
	\left( 2 \pi \right)^{-\frac{1}{2}}
	\text{det}^\frac{1}{4}\, C(\eta_0) \exp \left[ -\frac{N \lambda}{2 |g_2|}
	\vec{x}^T C(\eta_0) \vec{x} \right]
\end{align}
where $C(\eta_0) \equiv A(\eta_0)^{-1} \sqrt{\mathbb{1} + \frac{2
|g_2|}{\lambda} A(\eta_0)}$.  We can then project $\ee^{-S}\hat{\mathcal{H}}_\eta \ee^S$
into this low-energy subspace to obtain an effective Hamiltonian as
\begin{align*}
	\hat{\mathcal{H}}^{\rm eff}_\eta = \int d \vec{x} \; \psi^*_0(\vec{x}) \ee^{-S} \hat{\mathcal{H}}_\eta \ee^S
\psi_0 (\vec{x}).
\end{align*}

Now observe the following expectation value:
\begin{equation}
	M_{i j} \left\langle \partial_i \partial_j \right\rangle =
	-N \text{Tr} \left[ M \left( \mathbb{1} + \frac{2
	|g_2|}{\lambda} A(\eta_0) \right)^{-\frac{1}{2}} \right]
	\label{eq:obd-first-exp-values}
\end{equation}
where $M$ is an arbitrary matrix. Observe that this case covers the
coefficients of both the linear and quadratic $\delta\eta$ terms in
$\hat{\mathcal{H}}_\eta$, Eq.~(\ref{eq:zeroth-first-order-obd-hamiltonian}), by choosing
$M$ to be $-\frac{|g_2|}{2 N} B'(\eta_0)$ and $-\frac{|g_2|}{2 N}B''(\eta_0)$,
respectively. At this point note that should the expectation value of the
linear $\delta\eta$ term be of its natural order, order $1$, completing the
square in $\delta\eta$ would yield another term of order 1, invalidating its
placement into $\hat{\mathcal{H}}_\eta$ which is supposed to be of higher order in $1/N$.
Note also that the coefficient of the linear $\delta\eta$ term is exactly the
derivative of the zeroth-order energy $E_0(\eta_0)$ from
Eq.~(\ref{eq:obd-zeroth-order-gs-energy}) with respect to $\eta_0$. The above
problem is avoided if we expand about a local extremum of $E_0(\eta_0)$,
eliminating the linear term. For $\hat{\mathcal{H}}_\eta^\text{eff}$ to be bounded from
below, the extremum must be a minimum. Note that we do not get any apparent
order inconsistencies if we expand about an $\eta_0$ a distance of order
$N^{-1/4}$ away from the local minimum, but the analysis is vastly simplified
when the linear term is exactly zero, particularly for the last line of
Eq.~(\ref{eq:zeroth-first-order-obd-hamiltonian}) when close to $\eta_0 = 0$,
so we focus on expansions about zeroth-order energy minima from now on.
\begin{figure}
	\includegraphics[width=\columnwidth]{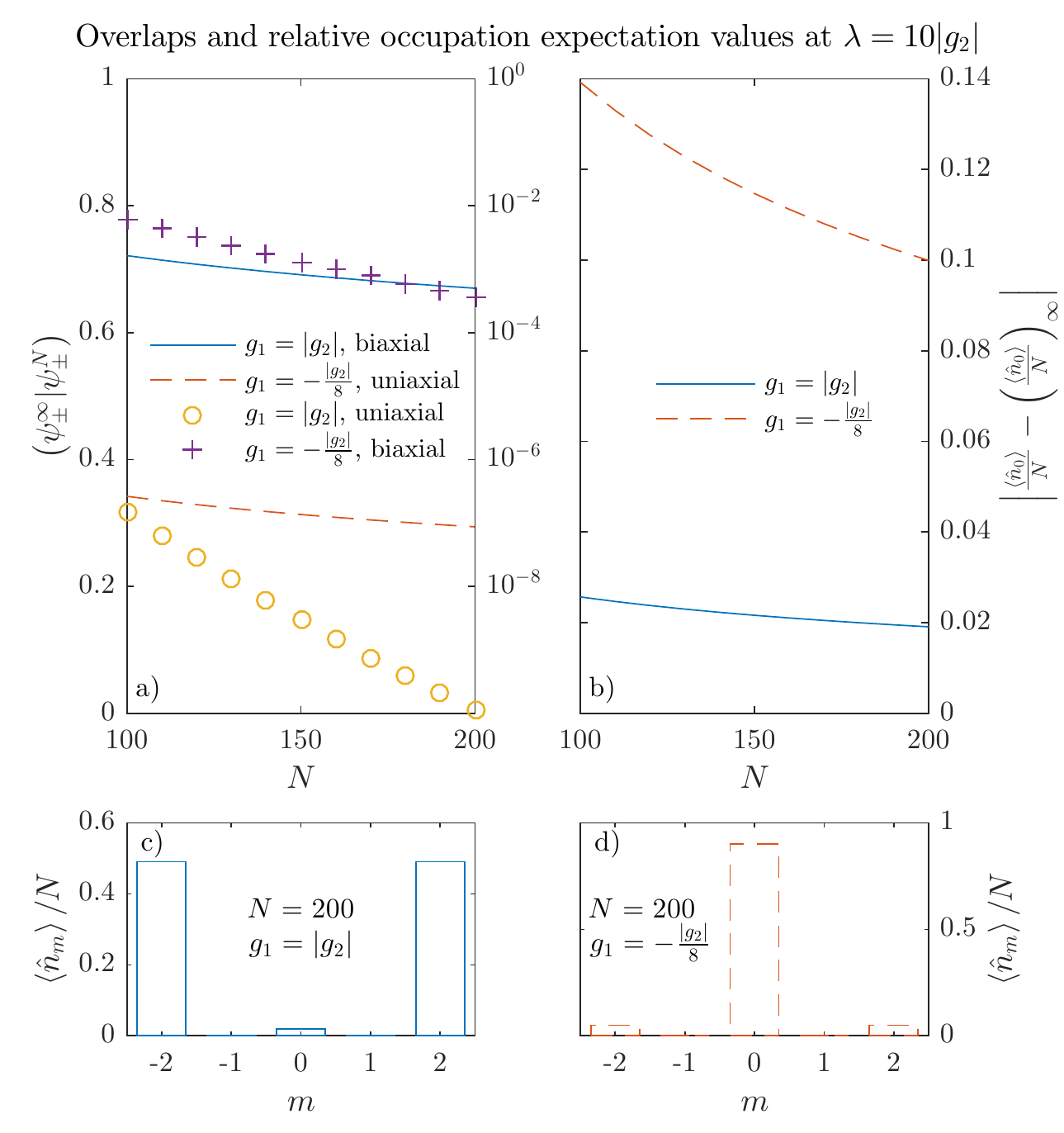}
	\caption{
	a) Absolute value of overlaps between the limiting states
	$\ket{\psi_\pm^\infty}$ and numerically computed ground states with
	respect to particle number $N$ at $\lambda = 10 |g_2|$.  The solid and
	dashed lines show the bigger overlaps, with the biaxial state for $g_1
	> 0$ and uniaxial for $g_1 < 0$, and correspond to the linear scale on
	the left y-axis. The markers show the smaller overlaps and correspond
	to the logarithmic scale on the right y-axis. We see that all
	mean-field overlaps decrease with particle number, in agreement with
	our analytical findings. b) The numerically computed expectation
	value of the fraction of particles in the $\hat{F}_z = 0$
	single-particle state with respect to particle number $N$ at $\lambda =
	10 |g_2|$. For compactness, the quantity actually plotted is $\left|
	\frac{\left\langle \hat{n}_0 \right\rangle}{N} - \left(
	\frac{\left\langle \hat{n}_0 \right\rangle}{N} \right)_\infty \right|$
	where $\hat{n}_m = \hat{a}_m^\dagger \hat{a}_m$, $\left\langle \cdot
	\right\rangle$ denotes the ground-state expectation value and $\left(
	\cdot \right)_\infty$ denotes taking the limit of $N \rightarrow
	\infty$.  $\left( \frac{\left\langle \hat{n}_0 \right\rangle}{N}
	\right)_\infty$ is predicted analytically and equals 0 for positive
	$g_1$ and 1 for negative $g_1$ (see main text). Note that
	$\left\langle \hat{n}_0 \right\rangle = N - \left\langle \hat{n}_2 + \hat{n}_{-2}
	\right\rangle$ to a very good approximation, with $\left\langle \hat{n}_1 +
	\hat{n}_{-1} \right\rangle$ already being negligible for the values of $N$
	shown.  The differences decrease with $N$, indicating a good agreement
	with analytical computations. These qualitative features are visible in
	c) and d) showing relative occupations of individual single-particle
	magnetic sublevels, labeled by $m$, at $N = 200$ for both signs of
	$g_1$. Results shown in c) and d) are obtained through exact diagonalization.}
	\label{fig:obd-figure}
\end{figure}

For large enough $\lambda$, these occur only at $\eta_0 = 0$ and $\pi/2$. In
both of these cases, $A(\eta_0)$ is proportional to $\mathbb{1}$, so both
$\psi_0(\vec{x})$ and $\ee^{\pm S}\psi_0(\vec{x})$ are isotropic in $x_1, x_2$.
As is easy to verify, this makes the expectation values of the last line of
Eq.~(\ref{eq:zeroth-first-order-obd-hamiltonian}) zero, eliminating those terms
from $\hat{\mathcal{H}}_\eta^\text{eff}$. Additionally $B'(\eta_0) \propto \text{diag}\,
\left( 1, -1 \right)$ which, combined with isotropy in $\vec{x}$, leads to
$B'_{i j}(\eta_0) \left\langle x_i \partial_j \right\rangle = 0$ as well.
Finally noting $B(0,\frac{\pi}{2}) = \pm \frac{g_1}{|g_2|} \mathbb{1}$ and
$B''(\eta_0) = -4 B(\eta_0)$ and evaluating the  coefficient of the quadratic
$\delta\eta$ term via Eq.~(\ref{eq:obd-first-exp-values}), we obtain
\begin{eqnarray}
	\hat{\mathcal{H}}_\eta^\text{eff} &=& -\frac{|g_2|}{2 N} \left( \partial_\eta^2 +
	\cot \eta\, \partial_\eta \right)\\
	&\mp& 4 g_1\left( 1 + 2 \left[ |g_2| + \left( 2 \pm 1 \right)g_1
	\right]/\lambda \right)^{-\frac{1}{2}} \delta\eta^2\nonumber
	\label{eq:obd-effective-delta-eta-hamiltonian}
\end{eqnarray}
where the upper sign corresponds to the expansion about $\eta_0 = 0$ and the
lower sign about $\eta_0 = \pi/2$. This immediately implies the ground-state is
localized about the pole, $\eta_0 = 0$, for negative $g_1$, and the 4-5
equator, $\eta_0 = \pi/2$, for positive $g_1$. In the latter case, we may
discard the $\cot \eta\,\partial_\eta$ term to obtain a 1-dimensional harmonic
oscillator Hamiltonian. Letting $d_{\frac{\pi}{2}} \equiv 1 + \frac{2}{\lambda}
\left( |g_2| + g_1 \right)$, we may write the effective mode energy and
oscillator length as $\Delta E_{\frac{\pi}{2}} = 2 \sqrt{\frac{2|g_2|g_1}{N}}
d_{\frac{\pi}{2}}^{-1/4}$ and $\xi_{\frac{\pi}{2}} = \left( \frac{|g_2|}{8 N
g_1} \right)^{1/4} d_{\frac{\pi}{2}}^{1/8}$. In the former case, when $g_1 <
0$, we may approximate $\cot \eta \simeq \eta^{-1}$, yielding a two-dimensional
isotropic harmonic oscillator Hamiltonian with the angular momentum term
absent. This may be solved by reintroducing the angular momentum term and then
restricting to isotropic, zero-angular-momentum states. Denoting $d_0 \equiv 1
+ \frac{2}{\lambda} \left( |g_2| + 3 g_1 \right)$, the effective spectrum
equals $E_0^n = (2 n + 1) \Delta E_0$ where $n = 0, 1, 2, \dots$ and $\Delta
E_0 = 2 \sqrt{\frac{2 |g_2| g_1}{N}} d_0^{-1/4}$, and the oscillator length, or
scale of localization in $\eta$, equals $\xi_0 = \left( \frac{|g_2|}{8 N g_1}
\right)^{1/4}d_0^{1/8}$.

In both cases, states are seen to be localized in the $\eta$ direction on the
scale of $\xi_\lambda \sim N^{-1/4}$. It may be verified by integration over
the 4-sphere, akin to the treatment in Sec.~\ref{sec:q-overlaps}, that this
causes the overlaps with \emph{any} mean-field state, i.e. a state of the form
of Eq.~(\ref{eq:rotor-basis-definition}), to tend to zero with increasing $N$.
While computationally accessible particle numbers are hardly in the large-$N$
regime, the numerical results in Fig.~\ref{fig:obd-figure} support our
analytical conclusions or, for the larger datasets, indicate the correct trend
with respect to $N$. Also shown in Fig.~\ref{fig:obd-figure} are numerical
results for the occupation numbers $\langle \hat{a}_n^\dagger \hat{a}_n
\rangle$ which verify the anaytical results of this section.

\section{Conclusion}
In this  work, we have developed and employed  the spin-2 rotor mapping
formalism to obtain a number of results that have so far proven analytically
inaccessible by other means. We have obtained an exact Hermitian Hamiltonian
for the special case of $g_1 = 0, g_2 < 0$ in the presence of an arbitrary
quadratic Zeeman field, and an approximate Hamiltonian in the $N|q| \gg
|g_{1,2}|$ regime for the entire nematic region.  Its spectrum and localization
width, the latter in the related $N \gg 1$ regime, were evaluated analytically
and found to be in good agreement with numerical results. Notably, for large
negative $q$ the ground state tends to a fragmented condensate, the excitations
about which cannot be analyzed by means of conventional Bogoliubov theory, but
do lend themselves to an analysis within the rotor framework.

Additionally, one finds that no traces remain of the order-by-disorder
mechanism, predicted to occur in the related continuum problem. The emergent
first-order phase transition at $g_1=0$ also seems to be gone and the behaviour
is smooth across the entire nematic region. Motivated by this, we considered an
alternative potential which leaves the mean-field degeneracy intact, and again
applied the rotor methodology. The ground state overlaps with \emph{all}
mean-field states are predicted to approach zero with increasing particle
number, indicating we are dealing with a highly non-mean-field state.  Its
individual magnetic sublevel occupation values are, however, consistent with
continuum order-by-disorder results.

The present analysis demonstrates that the rotor mapping may be fruitfully
applied to a number of different potentials for the tightly-confined spin-2 
problem. Additionally, simple analytical expressions may be obtained in the
relevant limits. This makes it a suitable candidate for application to further
specialized problems within the context of tightly confined spin-2 condensates.
An interesting avenue for further theoretical investigation of the mapping is
also the aforementioned Hermitianizing transform that may be applied in more
general setups. Preliminary analysis suggests that it is indeed applicable to
an arbitrary Hermitian bilinear term in the many-body Hamiltonian of
Eq.~(\ref{eq:many-body-hamiltonian}).

\acknowledgements
We thank Janne Ruostekoski for useful discussions during the early stages of this work.  This work was supported in part by the European Union's Seventh Framework Programme for research, technological development, and demonstration under Grant No.\  PCIG-GA-2013-631002.

\appendix
\section{Collective modes of continuum Hamiltonian under external field}

In this Appendix, we give the collective modes of the continuum
Hamiltonian described in  Sec.\  \ref{sec:background} under the external
potential given in Eq.\ (\ref{newfield}).  Obtaining the modes involves a
straightforward but lengthy Bogoliubov analysis.  Assuming we are in the
nematic region of the phase diagram, we insert $\vec{\hat{\psi}} ({\bf r})=
(\sin(\eta)/\sqrt{2},0,\cos(\eta),0,\sin(\eta)/\sqrt{2})^T \sqrt{\bar{\rho}} +
\vec{\hat{\phi}}({\bf r})$, where $\bar{\rho}$ is the constant mean-field
density, into the continuum Hamiltonian (with chemical potential) and expand to
quadratic order in $\vec{\hat{\phi}}({\bf r})$.  After diagonalizing the
resulting Hamiltonian, one finds the mode energies of the usual form 
\begin{align}
E_{{\bf k},n}= \sqrt{\xi_{{\bf k},n}(\xi_{{\bf k},n}+ 2 \gamma_n)}
\end{align}
where the particular parameters for the five modes are
\begin{align}
\xi_{{\bf k},1}&=\xi_{{\bf k},2}=\xi_{{\bf k},3}= \varepsilon_{\bf k} \\
\xi_{{\bf k},4}&= \xi_{{\bf k},5}= \varepsilon_{\bf k} + \lambda
\end{align}
and
\begin{align}
\gamma_1 &= - c_2 \bar{\rho}  \\
\gamma_2 &= (c_0 + c_2) \bar{\rho}  \\
\gamma_3 &=  (4 \sin^2(\eta)c_1-c_2) \bar{\rho}  \\
\gamma_4 &=  (4 \sin^2(\eta+2\pi/3)c_1-c_2) \bar{\rho}  \\
\gamma_5 &=  (4 \sin^2(\eta-2\pi/3)c_1-c_2) \bar{\rho}.
\end{align}
Here, $\varepsilon_{\bf k} = \frac{k^2}{2m}$ is the free particle
dispersion.  We next turn to an analysis of the zero-point energy due to these
modes, namely $\Delta E = \frac{1}{2} \sum_{{\bf k},n} (E_{{\bf k},n}-E_{{\bf
k},n}|_{\eta=0} )$ where $E_{{\bf k},n}|_{\eta=0}$ is subtracted to
regularize the summation. It is found that, for sufficiently large $\lambda>0$,
the biaxial nematic state is selected when $g_1>0$ while the uniaxial nematic
state is selected when $g_1<0$.  This is consistent with the rotor treatment in
the main text.

\bibliography{sources2}
\end{document}